\documentclass[12pt]{article}
\pdfoutput=1
\usepackage{tikz}
\usepackage{mathrsfs}
\usepackage{amsmath,amsfonts}
\usepackage{latexsym}
\usepackage{amsthm}
\usepackage{amssymb}
\usepackage{subfig}
\usepackage{xfrac}
\usepackage{enumerate}
\usepackage{adjustbox}
\usepackage{extarrows}
\usepackage{cite}
\definecolor{darkblue}{cmyk}{0.9,0.9,0,0}
\usepackage[colorlinks=true,linkcolor=darkblue,citecolor=darkblue,urlcolor=darkblue]{hyperref}
\usepackage{wasysym}
\usepackage{varioref}
\usepackage[english]{babel}
\usepackage{booktabs}

\usepackage[font={small}]{caption}

\usepackage{microtype}

\newcommand{\zb}{\bar{z}}

\usepackage{relsize}
\usepackage{multirow}
\usepackage{mathtools}
\usepackage{comment}
\usepackage{dsdshorthand}
\usepackage[utf8]{inputenc}
\newcommand{\bea}{\begin{equation}\begin{aligned}}
\newcommand{\eea}[1]{\label{#1}\end{aligned}\end{equation}}
\newcommand{\boa}{\begin{align}}
\newcommand{\eoa}{\end{align}}
\newcommand{\beq}{\begin{equation}}
\newcommand{\eeq}{\end{equation}}
\def\d{\delta}
\def\D{\Delta}

\def\f{\phi}

\def\a{\alpha}
\def\e{\epsilon}
\def\ep{\epsilon}
\def\zb{\bar{z}}
\newcommand{\nn}{\nonumber\\}

\DeclareMathOperator\Disc{Disc}

\usepackage{tikz}
\usetikzlibrary{arrows,calc,shapes,decorations.pathmorphing,decorations.markings}
\tikzset{
>=stealth',
help lines/.style={dashed, thick},
axis/.style={<->},
important line/.style={thick},
connection/.style={thick, dotted},
  cross/.style={
    cross out,
    draw=black, 
    minimum size=7pt, 
    inner sep=0pt,
    outer sep=0pt
  },
  branchcut/.style={
    decoration={
      snake,
      amplitude=1pt,
      segment length=6pt,
    },
    decorate,
    thick
  },
->-/.style={decoration={
  markings,
  mark=at position #1 with {\arrow{>}}},postaction={decorate}}
}

\usepackage{tensor}
\usepackage{tikz}
\usetikzlibrary{shapes, arrows.meta, positioning}

\begin{document}

\thispagestyle{empty}

\renewcommand{\thefootnote}{\fnsymbol{footnote}}
\setcounter{page}{1}
\setcounter{footnote}{0}
\setcounter{figure}{0}


\noindent

\hfill
\begin{minipage}[t]{35mm}
\begin{flushright}
UUITP-45/20 
\end{flushright}
\end{minipage}

\vspace{1.0cm}
\numberwithin{equation}{section}
\begin{center}
{\Large\textbf{\mathversion{bold}
Two applications of the analytic conformal bootstrap:  A quick tour guide\\
}\par}

\vspace{1.0cm}

\textrm{Agnese Bissi, Parijat Dey and Giulia Fardelli}
\\ \vspace{1.2cm}
\footnotesize{\textit{
Department of Physics and Astronomy, Uppsala University
Box 516, SE-751 20 Uppsala, Sweden
}  
\vspace{4mm}
}

\end{center}

\begin{abstract}
We review the recent developments in the study of conformal field theories in generic space time dimensions using the methods of the conformal bootstrap, in its analytic aspect. These techniques are based solely on symmetries, in particular in the analytic structure and in the associativity of the operator product expansion. We focus on two applications of the analytic conformal bootstrap: the study of the $\epsilon$ expansion of the Wilson Fisher model via the introduction of a dispersion relation and the large $N$ expansion of maximally supersymmetric Super Yang Mills theory in four dimensions. 

\end{abstract}
\noindent
\setcounter{page}{1}
\renewcommand{\thefootnote}{\arabic{footnote}}
\setcounter{footnote}{0}

\setcounter{tocdepth}{2}

\newpage

\parskip 5pt plus 1pt   \jot = 1.5ex

\newpage

\parskip 5pt plus 1pt   \jot = 1.5ex

\tableofcontents

\section{Introduction}
Conformal field theories (CFTs) are ubiquitous in theoretical physics, as~they play a crucial role in several setups spanning from statistical models and condensed matter physics to holographic theories. The~symmetry group associated to conformal transformations in $d$ space--time dimensions\footnote{In this note, we will mostly deal with $d>2$ dimensional conformal field theories.} highly constrains the structure of the observables in these theories, completely fixing the space--time structure of two- and three-point correlators up to a set of coefficients (the conformal dimensions and the so-called three-point function coefficients). In contrast to ordinary quantum field theories, conformal field theories are equipped with a convergent operator product expansion (OPE) whose radius of convergence is finite. This structure allows us to write the product of two fields sitting in positions close to each other, as~a linear combination of fields at a middle point. In~particular, when inserted inside correlation functions, the~OPE is particularly useful because it makes it possible to express $n$ point functions as a sum over $(n-1)$ point functions.  By~repetitively using the OPE, it is then possible to reduce any $n$ point function to a sum of two- and three-point functions. In~addition, the~OPE is associative and this property is crucial to obtain consistency conditions that constrain the two- and three-point coefficients, which is the set of quantities determining the dynamics of a~CFT.

This approach goes under the name of the conformal bootstrap. Despite the fact that its original formulation goes back to the 1970s~\cite{Polyakov:1974gs,Ferrara:1973yt}, a~more recent numerical approach revived the interest in it~\cite{Rattazzi:2008pe}. The~main idea is to use the associativity of the OPE inside four-point functions to be able to put numerical bounds on the conformal dimension and the three-point function coefficient (OPE data) of the lightest operator present in the OPE of the two operators appearing in the four-point function  we started with. Over~the years, these techniques proved to be extremely efficient and achieved impressive results, as can be seen in~\cite{Poland:2018epd} for a recent review. This progress motivates a complementary analytic study of the consistency conditions, which exploits the analytic structure of the equations together with information on the OPE structure and additional symmetries, when present. This approach gave a plethora of results, for~instance in the large spin sector~\cite{Alday:2016njk,Alday:2015eya} and in large $N$ theories~\cite{Heemskerk:2009pn,Aharony:2016dwx}.

In this note, we review mostly the latter, the~analytic approach. Despite their emergent simplicity, the~crossing relations are very intricate equations, and~in generic space--time dimensions, it is extremely complicated to systematically find solutions. Recently, this has been the focus of some investigations and it has become clear that an analytic approach can be developed to give powerful results. In~this
approach, it is possible to implement constraints that are more readily visible in Lorentzian rather
than Euclidean signature. Namely, by~focusing on a Lorentzian limit which selects the contribution
from operators with large spin, crossing symmetry predicts their dimensions and three-point function coefficients.
More precisely, in~Lorentzian signature, one can take a limit in which the external operators are null
separated. In~this limit, the correlator develops singularities which, by~crossing symmetry, are
mapped to the OPE data of large spin operators~\cite{Caron-Huot:2017vep,Alday:2016njk,Alday:2015eya}. The~knowledge of the singularities allows us to
compute the OPE data as an expansion in inverse powers of the spin. Crucially, this works on all
orders, in~practise allowing the full OPE data to be reconstructed just from singular terms.
This approach turns out to be very efficient, particularly in large $N$ theories where the
corresponding singularities can be systematically computed. In~particular, we discuss two applications of the analytic method: one based on the  usage of a dispersion relation and the second one mainly targeted towards the study of four-dimensional superconformal~theories. 

The structure of the paper is as follows. In Section \ref{sec:basics}, the~basics of conformal field theories are introduced, with~a focus mostly on conformal bootstrap techniques and their implications. In Section \ref{sec:disp}, a~dispersion relation for conformal field theory is introduced and the example of  the correlators in the Wilson--Fisher models in $d=4-\ep$ dimensions is discussed. In Section \ref{sec:SCFT}, we introduced the basics of superconformal field theories, mostly focusing on four-dimensional theories and on the classification of the operators. In \mbox{Section \ref{sec:N4}}, we report the case of the four-dimensional $\mathcal{N}$ = 4 super Yang--Mills theory and in particular, we present the methodology and the results to obtain the most transcendental piece of the graviton amplitudes in $AdS_5 \times S^5$.

\section{Basics of Conformal Field~Theory}
\label{sec:basics}
Conformal transformations are those transformations that locally preserve the angles between the curves. Under~a conformal transformation $x^{\mu} \rightarrow x'^{\mu}$ in a $d$-dimensional space ($\mu=1,2,\cdots d$), the metric tensor transforms as
\begin{align}\label{cftdef}
g'_{\mu \nu}(x') =\sigma(x) g_{\mu \nu}(x) 
\end{align}
where the function $\sigma(x)$ is known as the scale factor. 
Conformal transformations consist of the following infinitesimal~transformations:
\begin{itemize}
\item
Translation: $x^{\mu}\rightarrow x^{\mu}+a^{\mu}$
\item
Rotation: $x^{\mu}\rightarrow x^{\mu}+\omega^{\mu}_{\nu}\,x^{\nu}$
\item
Dilatation: $x^{\mu}\rightarrow \alpha\, x^{\mu } $ 
\item
Special conformal transformation (SCT): $x^{\mu}\rightarrow x^{\mu }+2x^{\mu} \,x\cdot b-b^{\mu} x^2 $ 
\end{itemize}
where $\omega_{\mu \nu}$ is an antisymmetric tensor and $a^{\mu},b^{\mu}$ are arbitrary~vectors.

The finite conformal transformations  corresponding to those infinitesimal ones along with the {generators are given by}
\begin{center}
\begin{tabular}{c c  }
 \underline{Transformation} & \underline{Generator}  \\ 
Translation: $x'^{\mu}= x^{\mu}+a^{\mu}$  &  $P_{\mu} =i \partial_{\mu}$  \\  
Rotation: $x'^{\mu}= \Lambda^{\mu}_{\nu}\, x^{\nu}$ & $M_{\mu \nu} =i (x_{\mu} \partial_{\nu}-x_{\nu} \partial_{\mu})$ \\   
Dilatation: $x'^{\mu}= \lambda \,x^{\mu } $ & $D=i x^{\mu} \partial_{\mu}$\\
SCT: $x'^{\mu}= \frac{x^{\mu}-(x.x)}{1-2 (b.x)+(b.b) (x.x)} $ & $K_{\mu} =i (2 x_{\mu} x^{\nu}\partial_{\nu}-x^2\partial_{\mu})$\,.
\end{tabular}\label{generators}
\end{center}

The generators form the conformal algebra, whose commutation relations in flat spacetime $g_{\mu \nu}(x)=\eta_{\mu \nu}$ are:
\begin{align}
\begin{aligned}\label{algebra}
[M_{\mu \nu}, P_{\alpha}]&=\eta_{\nu \alpha}P_{\mu}-\eta_{\mu \alpha}P_{\nu} \\
   [M_{\mu \nu}, K_{\alpha} ] &=\eta_{\nu \alpha}K_{\mu}-\eta_{\mu \alpha}K_{\nu} \\
 [M_{\mu \nu}, M_{\rho \sigma}]&=\eta_{\nu \rho} M_{\mu \sigma}-\eta_{\mu \rho} M_{\nu \sigma}+\eta_{\nu \sigma} M_{\rho \mu}-\eta_{\mu \sigma} M_{\rho \nu}\\
  [D, P_{\mu}]&=P_{\mu}\\
   [D,K_{\mu}]&=-K_{\mu}\\
   [K_{\mu},P_{\nu}]&= 2\eta_{\mu \nu} D-2M_{\mu \nu}\,.
   \end{aligned}
\end{align}

All the commutators that are not written above vanish. The~conformal group is  $SO(d+1,1) $ or $SO(d,2) $ in Euclidean or Lorentzian signature, respectively. 

Conformal field theory (CFT) is described by an infinite set of local operators specified by their scaling dimension $\D$ and {spin $\ell$  in the symmetric traceless representation ($\ell/2,\ell/2)$ }
for the OPE of two scalar operators.
It is customary to work with states that are eigenstates of the dilatation operator:
\begin{align}\label{eig}
D | \D \rangle =i \D | \D\rangle\,.
\end{align}

Here, $ | \D\rangle$ is a state that is created by the action of a local operator of scaling dimension $\D$ at the origin on the vacuum:
\begin{align}
| \D\rangle= O(0) | 0\rangle\,.
\end{align}

 The states  are in one-to-one correspondence with local operators in a CFT. Inserting a primary operator at the origin generates a state with scaling dimension $\D$  and this is the so-called {\it state-operator correspondence}. In~this article, we discuss unitary conformal field theories, which have bounds for the operator scaling dimension:
\begin{align}\label{ubound}
 \D \geq \begin{cases}
 d-2+\ell & {\rm{for }}\,\quad \ell > 0\, , \\
 \frac{d-2}{2} & {\rm{for }}\,\quad \ell = 0\, .
 \end{cases}
 \end{align}
 
In addition, the~action of $P_{\mu}$ and $K_{\mu}$ on the eigenstates of the dilatation generator increases and decreases the eigenvalue by unity, due to their mass dimensionality. If~we keep acting with $K_{\mu} $ on these states, we will eventually reach a state with negative scaling dimension. For~a unitary CFT, states with negative dimension are not allowed and we must have $K_{\mu} |\D\rangle =0$, after~acting with $K_{\mu}$ a finite number of times. The~operator that creates this state is called a primary operator of dimension $\D$. When acting $n$-times with $P_{\mu}$ on primary operators, it is possible to generate a tower of operators with dimension $\Delta+n$. These operators are called descendant~operators.  

In a CFT, the observables are the correlation functions of local operators. 
{We can write the product of two local operators $\f_i(x_1)$ and $\f_j(x_2)$ of scaling dimensions $\D_i$ and $\D_j$, respectively, as a sum over an infinite number of primary operators $\f_k$ of scaling dimension $\D_k$}.  This is known as the operator product expansion (OPE):
 {\begin{align}\label{ope}
\f_i(x_1) \f_j(x_2)=\sum_{k=0}^{\infty} \mathcal{C}_{ijk}(x_1- x_2, \partial_2) \f_k(x_2)\,,
\end{align}
where the coefficients $\mathcal{C}_{ijk}$ depend on the positions of the operators $\f_i, \f_j, \f_k$ as well as on their scaling dimensions and spins. This is a convergent expansion that is valid for the finite separation of the operators $x_1- x_2$. Conformal symmetry determines the coefficients $\mathcal{C}_{ijk}$  up to a numerical factor $\lambda_{ijk}$:
\begin{align}\label{ope2}
\mathcal{C}_{ijk}(x, \partial)=\lambda_{ijk} |x|^{\D_k-\D_i-\D_j}\left(1+\alpha x^{\mu}\partial_{\mu}+\beta x^{\mu}x^{\nu}\partial_{\mu}\partial_{\nu}+\sigma x^2 \partial^2+\cdots \right)
\end{align}
where  $\alpha, \beta, \sigma$ are numbers completely fixed by conformal symmetry. We quote here these numbers for the special case  $\D_i=\D_j$ and $\D_k=\D$ and $d$ space--time dimension:
\begin{align}
\alpha=\frac{1}{2}\,,\quad \beta= \frac{\D+2}{8 (\D+1)}\,,\quad \sigma=-\frac{\D}{16(\D-\frac{d-2}{2})(\D+1)}\,.
\end{align}

The coefficient $\lambda_{ijk} $ is known as the OPE coefficient. }

 The power of CFT lies in the fact that it fixes the one-, two- and three-point functions, up~to a set of coefficients $\D$ and  $\lambda_{ijk}$, with~a fixed space--time dependence:
\begin{align}
\langle \f(x) & \rangle =0 \,, \label{1ptfn}\\
\langle \f_i(x_1) \f_j(x_2) \rangle &=\frac{\d_{i j}}{x_{12}^{\D_i}}\,, \label{2ptfn}\\
\langle \f_i(x_1) \f_j(x_2) \f_k(x_3) \rangle &=\frac{\lambda_{ijk}}{x_{12}^{\D_1+\D_2-\D_3}x_{23}^{\D_2+\D_3-\D_1} x_{13}^{\D_1+\D_3-\D_2}}\, \label{3ptfn},
\end{align}
where $\D_i$ is the scaling dimension of $\f_i(x)$ and  $x_{ij}=x_i-x_j$. Note that the coefficients of the two-point function can be {re-absorbed} by redefining the fields and then the coefficients of the three-point functions cannot be further re-absorbed.
The three-point function is fixed up to a constant $\lambda_{ijk}$.
{To see that this $\lambda_{ijk}$ is the same number that appears in  \eqref{ope2}, we apply the OPE \eqref{ope} to a three-point function and use the fact that the two-point function is non vanishing only when both operators are the same \eqref{2ptfn}. This kills the sum in \eqref{ope} to one operator and we are left with the form \eqref{3ptfn}. }

The  numbers that specify a CFT, namely the spectrum or scaling dimensions  and the OPE coefficient of the operators, are known as the {CFT data}. If~we know all of these CFT data, then we can completely fix the theory.
 {The higher point correlation functions can be recursively computed reducing it to a lower point correlation function by using the OPE.}
 \mbox{In \eqref{1ptfn}--\eqref{3ptfn},} the~operators $\f_i$ are scalars ($\ell=0$), but~conformal symmetry fixes the correlators of spinning operators in a similar~way. 
 
Now let us consider the four-point function which is not fully fixed by conformal invariance and therefore encodes the dynamical information of the CFT. If~we consider four identical scalars of scaling dimension $\D_\f$, inserted at four different points, it is possible, using conformal symmetry, to~write the four-point function in terms of conformally invariant cross-ratios defined as
\begin{align}\label{crossrat}
u=z\,\bar{z}= \frac{x^2_{12}\,x^2_{34}}{x^2_{13}\,x^2_{24}}\,,\qquad v=(1-z)\,(1-\bar{z})=\frac{x^2_{14}\,x^2_{23}}{x^2_{13}\,x^2_{24}}\,.
\end{align}

 The four-point function takes the following form:
\begin{align}\label{4ptfn}
\langle \f(x_1) \f(x_2) \f(x_3 ) \f(x_4) \rangle& = \frac{1}{x_{13}^{2\D_\f}x_{24}^{2\D_\f}} F(z, \zb)\nn
&=\frac{1}{x_{13}^{2\D_\f}x_{24}^{2\D_\f}} (z\zb)^{-\D_\f}\sum_{\D, \ell} C_{\D,\ell} g^{(d)}_{\D, \ell}(z, \zb)
\end{align}
where $\D$ and $\ell$ denote the scaling dimension and spin of the operators $O$ being exchanged. The coefficients $C_{\D,\ell}$ are the square of the OPE coefficients $\lambda^2_{\phi \f O}$. We will use the term OPE coefficient for $C_{\D,\ell}$ in what follows.
The  function $g^{(d)}_{\D, \ell}(z,\zb)$ contains the contribution of a primary operator of dimension $\D$ and all of its descendants. These are the conformal blocks whose form is completely fixed by conformal symmetry. They satisfy a differential equation derived from the conformal Casimir~\cite{Dolan:2003hv, Dolan:2011dv}. The~conformal blocks are generally complicated functions of cross-ratios and the explicit representation is known as an integral representation. In~even space-time dimension, some closed form expressions are known. We quote the form of the conformal blocks in four dimensions below:
\begin{align}\label{cb}
g^{(4)}_{\D, \ell}(z,\zb) =\frac{z \zb}{z-\zb}\left(k_{\frac{\D+\ell}{2}}(z)k_{\frac{\D-\ell-2}{2}}(\zb)-k_{\frac{\D+\ell}{2}}(\zb)k_{\frac{\D-\ell-2}{2}}(z)\right)\, ,
\end{align}
where:
\begin{align}\label{block}
k_{\b}(z)=z^{\b}\,  _2F_1(\b, \b, 2\b,z)\,.
\end{align}

Here, $_2F_1$ is a Gauss hypergeometric function.
Inside the four-point function \eqref{4ptfn}, {we can fuse together operators} ({12}) and ({34}). This is the $s$-channel expansion of the correlator. We could have also expanded into the $t$-channel where we fuse  ({14}) and ({23}), or~in the $u$-channel where we fuse ({13}) and ({24}). Since the OPE is associative, these three expansions must be the same. This is the statement of crossing symmetry which is fully equivalent to the associativity of the OPE, and~results in the following equation: 
\begin{align}\label{booteq}
F(z,\zb)= F(1-z,1-\zb)= (z\zb)^{-\D_\f}F(1/z, 1/\zb)\,.
\end{align}

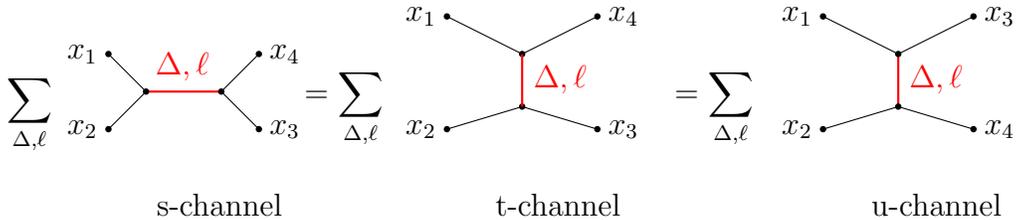
\begin{figure}[h]
\centering
\begin{tikzpicture}[scale=1]
\draw[red,thick]  (1,0.5) -- (0,0.5) node [text width=2 cm,midway,above,align=center ] {$\D, \ell$} ;
\filldraw[black] (1,0.5) circle (1pt) ;
\filldraw[black] (0,0.5) circle (1pt) ;
\draw   (0,0.5) -- (-0.5,1) node[left] {$x_1$};
\filldraw[black] (-0.5,1) circle (1pt) ;
\draw  (0,0.5)-- (-0.5,0)node[left ] {$x_2$} ;
\filldraw[black] (-0.5,0) circle (1pt) ;
\draw   (1,0.5) -- (1.5,1) node[right]{$x_4$};
\filldraw[black] (1.5,1) circle (1pt) ;
\draw   (1,0.5) -- (1.5,0) node[right]{$x_3$};
\filldraw[black] (1.5,0) circle (1pt) ;
\draw (0,-1)  node[anchor=west] {s-channel};
\draw (4.5,-1)  node[anchor=west] {t-channel};
\draw (9.5,-1)  node[anchor=west] {u-channel};
\draw (-2,0.2)  node[anchor=west] {$\underset{{\Delta,\ell}}{\mathlarger\sum}$};
\put(60,10){$= \underset{{\Delta,\ell}}{\mathlarger\sum}$};
\filldraw[black] (5,1) circle (1pt) ;
\filldraw[black] (5,0.3) circle (1pt) ;
\filldraw[black] (4,0) circle (1pt) ;
\filldraw[black] (6,0) circle (1pt) ;
\draw  (5,0.3)-- (6,0) node[right]{$x_3$};
\draw   (5,0.3)-- (4,0)node[left ] {$x_2$} ;
\draw[red,thick]  (5,0.3)-- (5,1) node[anchor=north west] {$\Delta,\ell$};
\filldraw[black] (6,1.5) circle (1pt) ;
\draw   (5,1)-- (6,1.5) node[right]{$x_4$};;
\filldraw[black] (4,1.5) circle (1pt) ;
\draw    (5,1) -- (4,1.5) node[left ] {$x_1$};
\put(200,10){$=\underset{{\Delta,\ell}}{\mathlarger\sum}$};
\draw[red,thick]  (10,0.3)-- (10,1) node[anchor=north west] {$\Delta,\ell$};
\filldraw[black] (10,1) circle (1pt) ;
\filldraw[black] (10,0.3) circle (1pt) ;
\filldraw[black] (9,0) circle (1pt) ;
\filldraw[black] (11,0) circle (1pt) ;
\draw  (10,0.3)-- (11,0) node[right]{$x_4$};
\draw  (10,0.3)-- (9,0) node[left]{$x_2$};
\filldraw[black] (9,1.5) circle (1pt) ;
\draw  (10,1)-- (9,1.5) node[left]{$x_1$};
\filldraw[black] (11,1.5) circle (1pt) ;
\draw  (10,1)-- (11,1.5) node[right]{$x_3$};
\end{tikzpicture}
\caption{Crossing symmetry as a different expansion in the three~channels.  }\label{fig:crossing} 
\end{figure}

This is the conformal bootstrap equation. The~crossing symmetry is depicted in \mbox{Figure~\ref{fig:crossing}}. The conformal bootstrap is a self-sustaining process that is supposed to continue without any external input and entirely relies upon the symmetry of the CFT. We focus on the CFT itself without worrying about a specific microscopic realisation and this is a Lagrangian free approach. \eqref{booteq} is a functional constraint on the CFT data and must be satisfied for all values of the cross-ratios $z, \zb$. However, this is a complicated constraint as it involves a double infinite sums over the operator spectrum and spin. It is not possible to generically solve this equation analytically and extract the CFT data. There are several approaches to extract the CFT data by solving \eqref{booteq}, both analytical and numerical. One efficient method is the numerical one, which is a numerical procedure that allows finding bounds on the CFT data for the operators appearing in the OPE decomposition, by~using the relation \eqref{booteq} and other symmetries that the theory may possess, as can be seen for instance in a recent review~\cite{Poland:2018epd}. 
In the next sections, we discuss some of the analytic methods to study the same relations~\cite{Bissi:2019kkx}.

\section{Dispersion Relation in~CFT}
\label{sec:disp}
In this section, we present a dispersion relation for the CFT four-point correlation function following~\cite{Bissi:2019kkx}. The dispersion relation makes it possible to construct a function from the knowledge of its discontinuity. We will exploit the analytic properties together with the crossing symmetry of the correlator \eqref{booteq} and show that in perturbative CFT, where we have an expansion of the CFT data in a perturbative parameter, the~four-point function only depends on the spectrum of the theory and the OPE coefficients of certain low lying operators\footnote{In this context, the~meaning of \textit{low lying} refers to the dimension of the operators in the OPE.}. 

\subsection{Analytic Structure of Conformal~Blocks}
Let us begin by analysing the analytic structure of the conformal blocks   in $d$ dimensions~\cite{Dolan:2000ut}:
\beq\label{gd}
g^{(d)}_{\De,\ell}(z,\zb) = (z \zb)^{\frac{\De-\ell}{2}} {\tilde g}^{(d)}_{\De,\ell}(z,\zb)\,,
\eeq
where:
\beq\label{gtilde}
{\tilde g}^{(d)}_{\De,0}(z,\zb)=\sum\limits_{m,n=0}^\infty
\frac{\left(\frac{\De}{2} \right)^2_m 
\left(\frac{\De}{2} \right)^2_{m+n}  }
{m! n! (\De+1-\frac{d}{2})_m (\De)_{2m+n}} z^m \zb^m (z+\zb -z \zb)^n\,.
\eeq
is the conformal block for the scalar exchange operators. The~conformal blocks for the exchange of spinning operators can be obtained from the scalar blocks by a recursion relation~\cite{Dolan:2000ut}. The~sum over $n$ in  \eqref{gtilde}   results in a hypergeometric function. One can further use the Euler integral representation for the hypergeometric function and rewrite  \eqref{gtilde}    in the following form: \beq\label{gtildeint}
{\tilde g}^{(d)}_{\De,0}(z,\zb) 
= \frac{\Gamma(\De)}{\Gamma^2(\frac{\De}{2}) } \int\limits_0^1 dt
\frac{{}_2 F_1 \left(\tfrac{\De}{2},\tfrac{\De}{2},1-\tfrac{d}{2}+\De, \tfrac{t(1-t)z \zb}{1-t(z+\zb-z \zb)} \right)}{
t^{1-\frac{\De}{2}}
(1-t)^{1-\frac{\De}{2}}
(1-t(z+\zb-z \zb))^{\frac{\De}{2}}
}\,.
\eeq

It was shown in~\cite{Pappadopulo:2012jk} that \eqref{gtildeint} is analytic when:
\beq
z, \bar{z} \in \mathbb{C} \setminus (1, +\infty) \quad \text{with} \quad (1-z)(1-\zb) \in \mathbb{C} \setminus  (-\infty,0)\,.
\eeq

In what follows, we will study the analyticity in the variable $z$ and keep $\zb$  fixed to some value between $0$ and $1$.  Note  that  $0 < \zb <1$ lies on the $u$-channel branch cut which is on the boundary of the convergence region of the $u$-channel. It was shown in~\cite{Kravchuk:2020scc} that the OPE converges in this regime in a distributional sense. Hence, the domain of analyticity in $z$ becomes:
\beq
z \in \mathbb{C} \setminus (1, +\infty)\,.
\eeq

The conformal blocks for the exchange of spinning operators inherit the same analytic properties as they are given in terms of  \eqref{gtilde} by a recursion relation in $\ell$. The~conformal blocks have this specific structure in any space--time dimension.

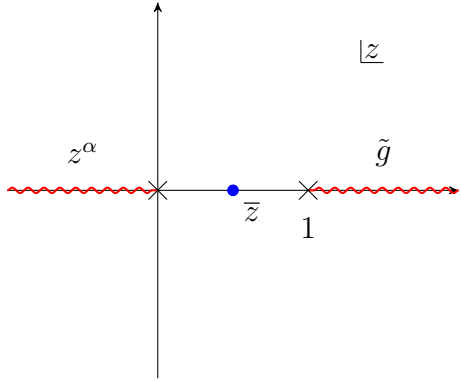
\begin{figure}[h] \centering
  \begin{tikzpicture}[scale=1]
    \coordinate (n) at (-1,2.5);
    \coordinate (e) at (3,0);
    \coordinate (w) at (-3,0);
    \coordinate (s) at (-1,-2.5);
    \coordinate (bp1) at (1,0);
    \coordinate (bp2) at (-1,0);
    \draw[->] (w) --  (e) ;
    \draw[->] (s) --  (n) ;
    \filldraw [blue] 
     (0,0) circle (2pt) node[below right, black] {$\zb$};
	\draw [red, branchcut] (w) -- (bp2);
	\draw [red, branchcut] (e) -- (bp1);
	\node at (1.85,1.85) [] {$ z$};
	\draw[-] (2,1.7) -- (1.7,1.7);
    \draw[-] (1.7,1.7) -- (1.7,2);
    \node at (1,0) [cross] {};
    \node at (-1,0) [cross] {};
    \node at (1,-0.5) [] {$1$};
    \node at (-2,0.5) [] {$z^{\a}$};
    \node at (2,0.5) [] {$\tilde{g}$};
  \end{tikzpicture}
\caption{Analytic structure of the conformal blocks from power (\textbf{left}) and hypergeometric function (\textbf{right}).} \label{fig1}
\end{figure}

The conformal blocks \eqref{gd} have a branch cut for $z < 0$ that originates from the non-integer powers of $z$. The~blocks have another branch cut for $z > 1$ originating from $\tilde{g}$. The~analytic structure of the conformal block is depicted in Figure~\ref{fig1}. Note that the discontinuity of the branch cut due to the overall power in the second line of \eqref{4ptfn} is much simpler, which results in:  
\beq\label{discsimp}
\underset{z\, <\, 0}{\Disc}\, (z \zb)^{-\De_\f}   g^{(d)}_{\De,\ell}(z,\bar{z}) = -2i{\sin\pi ( \De_\phi- \frac{\De- \ell}{2})}   (|z| \zb)^{-\De_\f+\frac{\De- \ell}{2}}{\tilde g}^{(d)}_{\De,\ell}(z,\bar{z})\,,
\eeq
where we define the discontinuity of a function $f(z)$ as
\beq
\underset{z}{\Disc}\, f(z) \equiv \lim_{\a \to 0^+} f(z + i \a) - f(z - i \a) \,.
\eeq

We will see how  \eqref{discsimp} plays  a key role in the CFT dispersion relation in the next~sections.

\subsection{Crossing Symmetry and Dispersion~Relation}
In this section, we exploit the analytic properties and crossing symmetry of the conformal correlator to present a dispersion relation. As~a first step, we introduce a pole at a generic point $z'$ and write the correlator \eqref{4ptfn} as the residue of this pole using Cauchy's residue theorem: 
\beq\label{disprelation_ansatz}
F(z,\zb) = \frac{1}{2 \pi i} \oint\limits_{z} dz' \, \frac{1}{z' - z} F(z',\zb)\,.
\eeq

\begin{figure}[h] \centering
  \begin{tikzpicture}[scale=1.5]
    \coordinate (n) at (-0,2);
    \coordinate (e) at (2,0);
    \coordinate (w) at (-2,0);
    \coordinate (s) at (-0,-2);
    \coordinate (bp1) at (1,0);
    \coordinate (bp2) at (0,0);
    \draw[->] (w) --  (e) ;
    \draw[->] (s) --  (n) ;
    \filldraw [blue] (0.33,0) circle (1pt);
    \filldraw [blue] (0.67,0) circle (1pt);
	\draw[red] [branchcut] (w) -- (bp2);
	\draw[red] [branchcut] (1.9,0) -- (bp1);
    \node at (1,0) [cross] {};
    \node at (-0,0) [cross] {};
    \node at (0.33,-0.25) [] {$z$};
    \node at (0.67,-0.25) [] {$\zb$};
    \node at (1,-0.25) [] {$1$};
    \node at (1.85,1.85) [] {$z'$};
    \draw[-] (2,1.7) -- (1.7,1.7);
    \draw[-] (1.7,1.7) -- (1.7,2);
    \draw[->-=.33] (2,0.1) arc (0:180:2);
    \draw[->-=.5] (-2,0.1) -- (0,0.1);
    \draw[-] (0,0.1) arc (90:-90:0.1);
    \draw[->-=.5] (0,-0.1) -- (-2,-0.1);
    \draw[->-=.33] (-2,-0.1) arc (180:360:2);
    \draw[->-=.5] (2,-0.1) -- (1,-0.1);
    \draw[-] (1,-0.1) arc (270:90:0.1);
    \draw[->-=.5] (1,0.1) -- (2,0.1);
    \draw[->-=.5] (0.33,-0.1) arc (-90:270:0.1);
  \end{tikzpicture}
\caption{Contour deformation for the dispersion~relation.} \label{fig2}
\end{figure}
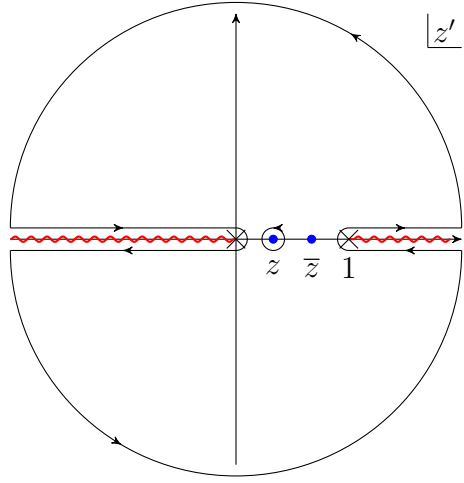

The analytic structure of this correlator follows from the analytic structure of the conformal blocks as discussed in the previous section. Now, we deform the contour to wrap around the branch cuts on the real axis as shown in Figure~\ref{fig2}. In~order to determine the arc contribution at $z=\infty$, we have to consider the Laurent series expansion of $F(z, \zb)$, which is given by the $u$-channel OPE: 
\beq
F(z,\bar{z}) =\sum\limits_{\De,\ell} C_{\De,\ell} g^{(d)}_{\De,\ell}(1/z,1/\bar{z})\,, \qquad
g^{(d)}_{\De,\ell}(z,\bar{z}) = (z \bar{z})^{\frac{\De-\ell}{2}} \left( \de_{\ell, 0} + O(z) +O(\zb)  \right)\,.
\eeq

Note that due to the overall prefactor $z^{\frac{\D-\ell}{2}}$, only primary operators with $\D-\ell \leq 0$ contribute to the arc at infinity. However, for~unitary CFTs in $d> 2$, the unitary bound \eqref{ubound} shows that identity ($\D=0,\ell=0$) is the only such operator that contributes to the arc and its arc {contribution is 1}\footnote{Note that this does not work for  $d=2$ correlators where there is no gap in the spectrum.}. This results in the following dispersion relation:
\beq\label{eq:dispersion_relation}
F(z,\zb) = 1 + \frac{1}{2 \pi i} \int\limits_{-\infty}^\infty dz' \, \frac{1}{z' - z} \underset{z'}{\Disc}\, F(z',\zb)\,.
\eeq
Now, we use the crossing symmetry \eqref{booteq} to express the discontinuity at $z >1$ in terms of the simpler discontinuity at $z < 0$: 
\beq
\underset{z\, >\, 1}{\Disc}\, F(z,\zb) 
= - \underset{z\, <\, 0}{\Disc}\, F(1-z,1-\zb) \Big|_{\substack{z\to 1-z\\\zb \to 1-\zb}} 
=- \underset{z\, <\, 0}{\Disc}\, F(z,\zb) \Big|_{\substack{z\to 1-z\\\zb \to 1-\zb}} \,.
\eeq
This can be used to rewrite the integral on the positive real axis in  \eqref{eq:dispersion_relation}  in terms of an integral on the negative real axis:
\beq
\int\limits_{1}^\infty dz' \, \frac{1}{z' - z} \underset{z'}{\Disc}\, F(z',\zb)
= \int\limits_{-\infty}^0 dz' \, \frac{1}{z' - (1-z)} \underset{z'}{\Disc}\, F(z',1-\zb)\,,
\eeq
Putting these together, we obtain the following dispersion relation:
\beq\label{dispfinal}
F(z,\zb) = 1 + \left(\frac{1}{2 \pi i} \int\limits_{-\infty}^0 dz' \, \frac{1}{z' - z} \underset{z'}{\Disc}\, F(z',\zb) + (z,\zb) \to (1-z,1-\zb)\right) \,.
\eeq
This shows that the correlator in a unitary CFT is determined by its discontinuity at $z<0$ together with crossing~symmetry.

Let us see how the dispersion relation can be applied to compute correlators in a mean field theory in $d$ dimensions. We assume that the identity operator is present in the OPE of the operators. Now, we show how the identity operator in the $s$ channel reproduces the mean field theory correlator. The~$s$ channel identity is given by $\frac{1}{(z \zb)^{\De_\f}}$, whose \mbox{discontinuity is:}
\begin{align}\label{discphase2}
\underset{z\, <\, 0}{\Disc}\, (z \zb)^{-\De_\f}   = -2i{\sin\pi  \De_\phi} (|z| \zb)^{-\De_\f}\,.
\end{align}

Using \eqref{dispfinal}, we obtain the following mean field theory correlator:
\begin{align}\label{mft0}
F^{MF}(z,\zb) = 1 + \frac{1}{(z \zb)^{\De_\f}} + \frac{1}{((1-z)(1- \zb))^{\De_\f}}\,.
\end{align}

{Note that plugging \eqref{discphase2} into  \eqref{dispfinal} will yield a finite result for $0< \Delta_\phi<1$. For~$\Delta_\phi \in \N$, the prefactor in \eqref{discphase2} vanishes. The~discontinuity in that case, coming from a pole, is a delta function and can be thought of as a distribution around $z=0$ which finally yields the same result as in \eqref{mft0}. However, one needs to analytically continue $\Delta_\phi$ in order to obtain \eqref{mft0} for generic values of $\Delta_\phi$. This analytic continuation is justified in perturbation theory where we  have analytic control. The~way to obtain the analytic continuation in general is to consider a subtracted dispersion relation which is not what we did. In~that sense, the use of this dispersion relation is limited.}

Now, we can decompose \eqref{mft0} into conformal blocks:
\begin{align}
F^{MF}(z,\zb) = \frac{1}{\left( z \zb \right)^{\De_\f}} \left( 1+ \sum\limits_{n=0}^{\infty} \sum\limits_{\substack{\ell=0}}^{\infty} C^{MF}_{n,\ell} \, g^{(d)}_{2 \De_\f +2n+\ell,\ell}(z,\zb)\right)\,,
\end{align}
to reproduce the mean field theory OPE coefficients~\cite{Fitzpatrick:2011dm}:
\begin{align}\label{mftope}
C^{MF}_{n,\ell} = \frac{(1+(-1)^\ell) (\De_\f-\frac{d}{2}+1)_n^2 (\De_\f)_{n+\ell}^2}{\ell!n! (\ell+\frac{d}{2})_n (2\De_\f+n-d+1)_n (2\De_\f+2n+\ell-1)_\ell (2\De_\f+n+\ell-\frac{d}{2})_n}\,.
\end{align}

Note that only even spin operators appear in the OPE when we study the correlator of four identical scalar~operators.

We will  now study perturbative CFT, where we have an expansion around the mean field theory in the  perturbative parameter $\e$. The~exchanged operators in the operator product expansion of $\f \times \f $ contain double trace operators of the schematic form $[\f \f]_{n, \ell}\sim \f \Box^{n}\partial^{\ell}\f$ with bare dimension $2 \D_\phi +2n+ \ell $.
We begin by expanding the CFT data as follows:
\begin{align}
\D_{n,\ell} &= 2 \D_\phi +2n+ \ell + \ep \gamma^{(1)}_{n,\ell}  + \ep^2 \gamma^{(2)}_{n,\ell}+O(\e^3)\,,\nn
C_{n,\ell} &= C_{n,\ell}^{MF} + \ep C_{n,\ell}^{(1)} + \ep^2 C_{n,\ell}^{(2)}+O(\e^3)\,.
\end{align}

This results in the following expansion of the correlator:
\beq
F(z,\zb) = F^{MF} (z,\zb) + \e F^{(1)} (z,\zb)  + \e^2 F^{(2)} (z,\zb)+O(\e^3)\,.
\eeq

The leading order correction to the correlator is given by
\begin{align}\label{corr1}
F^{(1)} (z,\zb) &= (z \zb)^{-\De_\f} \sum\limits_{n=0}^\infty \sum\limits_{\substack{\ell=0\\\text{even}}}^\infty 
\left( C_{n,\ell}^{(1)} + C_{n,\ell}^{(0)} \gamma^{(1)}_{n,\ell} \partial_{\e} \right) \, g^{(d)}_{2\De_\f+2n+\ell,\ell}(z,\bar{z})\,,
\end{align}
{where $\partial_{\e}g^{(d)}_{2\De_\f+2n+\ell,\ell}(z,\bar{z})$ is the derivative of  $g^{(d)}_{2\De_\f+2n+\ell,\ell}(z,\bar{z})$ with respect to $\e$.}
We compute the discontinuity of \eqref{corr1} at $z< 0$ using \eqref{discsimp}:
\beq\label{mft2}
\underset{z\, <\, 0}{\Disc}\, F^{(1)} (z,\zb) = \pi i (z \zb)^{-\De_\f} \sum\limits_{n=0}^\infty \sum\limits_{\substack{\ell=0\\\text{even}}}^\infty 
C_{n,\ell}^{(0)} \gamma^{(1)}_{n,\ell}  \, g^{(d)}_{2\De_\f+2n+\ell,\ell}(z,\bar{z})\,.
\eeq

Using \eqref{dispfinal}, we can compute the correlator $F^{(1)} (z,\zb)$. Since \eqref{mft2} determines the correlator from the dispersion relation, it follows that the CFT correlator in perturbative settings is entirely determined by the spectrum $\gamma^{(1)}_{n,\ell}$ at that order and the OPE coefficient $C_{n,\ell}^{(0)}$ at the previous order in the CFT. As~a next step we can decompose the correlator into conformal blocks and extract the OPE coefficient $C_{n,\ell}^{(1)}$. This process is summarised in Table \ref{tabledisp}.
\begin{table}
\begin{center}
\begin{tabular}{|c|c|l|l|l|}
 \hline
 \text{{\bf Input}}  &\text{{\bf Output} (Correlator) }  & \text{OPE data }   \\ \hline
   \text{Identity}: {$\Delta=0, \ell=0$}&$F^{(0)}(z, \bar{z})$ & $C^{MF}_{n, \ell}$  \\
\hline  $C^{MF}_{n, \ell},\quad \gamma^{(1)}_{n, \ell}$   &     $F^{(1)}(z, \bar{z})$  & $C^{(1)}_{n, \ell}$  \\
\hline    $C^{MF}_{n, \ell},\quad C^{(1)}_{n, \ell},\quad \gamma^{(1)}_{n, \ell}, \quad \gamma^{(2)}_{n, \ell}$   &$F^{(2)}(z, \bar{z})$  &$C^{(2)}_{n, \ell}$  \\
    \hline $\cdots$ &  $\cdots$ & $\cdots$\\
        \hline      
\end{tabular}
\caption{Input and Output for the dispersion relation}\label{tabledisp}
\end{center} 
\end{table}

\subsection{Computing Wilson--Fisher Correlator Using Dispersion~Relation}

In this section, we discuss how the dispersion relation \eqref{dispfinal} can be applied to compute the four-point correlation function in Wilson--Fisher  $\f^4$ theory in $d=4-\ep$ dimensions as a perturbative expansion in $\ep$. The~Wilson--Fisher theory is described by the Lagrangian:
\begin{align}
S=\int d^dx \left(\frac{1}{2}(\partial\f)^2+\frac{g \mu^{\e}}{4!}\f^4\right)
\end{align}
where $\mu$ is the energy scale and we have a fixed point for the coupling $g^{*}=\frac{16\pi^2}{3}\e+O(\e)$. The~correlator in the theory can be computed by perturbatively evaluating Feynman diagrams in the $\e$-expansion. Here, we will compute the correlator $\langle \f \f \f \f \rangle$ \mbox{using \eqref{dispfinal}}.
The correlation function and the CFT data contain the same amount of information. However, it may not always be possible to resum the CFT data and obtain a closed form expression for the correlator. It is easier to extract the CFT data from the closed form expression of the correlator. We will see that the dispersion relation allows us to directly compute the correlator without resumming the CFT data. In~perturbative CFT, this can be thought of as an alternative way of computing the correlator using the inputs (spectrum) from the Feynman diagrams. The~inputs we need here can be obtained from the two-point function which is much simpler to compute  than the four-point~function.

We expand the CFT data in $\e$ using the input from Wilson--Fisher theory:
\begin{align}
\De_\f &= 1 - \frac{1}{2} \e +\frac{1}{108} \e^2+O(\e^3)\,,\\
\De_{0} &= 2 \De_\f + \frac{1}{3} \e + \frac{8}{81} \e^2 +O(\e^3)\,,\\
\De_{\ell} &= 2 \De_\f + \ell - \frac{1}{9 \ell (\ell+1)} \e^2+O(\e^3)\, \qquad \ell >0\,.
\end{align}

From \eqref{mftope}, it is evident that only operators $\f \partial_{\mu_1}\cdots \partial_{\mu_{\ell}} \f$ appear in the OPE up to $O(\e)$:
\begin{align}
C^{MF}_{n, \ell}=\frac{(1+(-1)^{\ell}){(\D_\f)_{\ell}}^2}{\ell! (2\D_\f+\ell-1)_{\ell}}\delta_{n,0}+O(\e^2)\,.
\end{align}

 Since the anomalous dimensions of $\ell >0$ operators start at $O(\e^2)$, only the $\ell=0$ operator will contribute to the discontinuity of the correlator at $O(\e)$. The~associated discontinuity from $\D_0$ is given by
\begin{align}
\underset{z\, <\, 0}{\Disc}\, F^{(1)} (z,\zb) = \frac{2}{3} \pi i (z \zb)^{-1} g_{2,0}(z,\bar{z}) = 
\frac{2}{3}  \pi i \frac{\log(1-\zb) - \log(1-z)}{z-\zb}\,.
\end{align}

Using \eqref{dispfinal}, we obtain:
\begin{align}\label{f1}
F^{(1)} (z,\zb) = \frac{1}{3(z-\bar{z})}\left(\log(z \bar{z}) \log \left( \frac{1-\zb}{1-z}\right) -2 \text{Li}_2 (z)+2 \text{Li}_2 (\bar{z}) \right)\,.
\end{align}

\begin{figure} \centering
\begin{tikzpicture}[scale=1]
\draw node[left ] {$x_2$} (0,0) -- (1,1) node[right ] {$x_3$} ;
\draw   (1,0)node[right]{$x_4$} -- (0,1) node[left] {$x_1$};
\filldraw[black] (0,0) circle (1pt) ;
\filldraw[black] (1,0) circle (1pt) ;
\filldraw[black] (0,1) circle (1pt) ;
\filldraw[black] (1,1) circle (1pt) ;
\filldraw[black] (0.5,0.5) circle (1pt) ;
\end{tikzpicture}
\caption{Contact diagram at $O(\e)$.}\label{fig:contactdiagram} 
\end{figure}
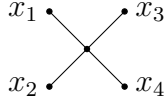

This corresponds to the contact diagram Figure~\ref{fig:contactdiagram}. 
The correlator \eqref{f1} can be decomposed into conformal blocks to obtain the following OPE coefficient:
\begin{align}
C_{n,\ell}^{(1)} = -\frac{2}{3} \de_{n,0} \de_{\ell,0}\,.
\end{align}

Now, we proceed to compute the correlator at the next order. Since we are only interested in the terms having discontinuity at $z<0$ we expand the correlator as follows:
\begin{align}
F(z, \zb) ={}& C_{0} \left( \frac{1}{2}(\gamma^{(1)}_{0} \e + \gamma^{(2)}_{0} \e^2) \log(z \zb) (1+\e \partial_\e) + \frac{1}{8} (\gamma^{(1)}_{0})^2 \e^2 \log(z \zb)^2 \right) {\tilde g}^{(4-\e)}_{\De_{0},0} (z,\zb)\\
&+ \sum\limits_{\substack{\ell=2\\\text{even}}}^\infty 
\frac{1}{2} C_{\ell} \gamma^{(2)}_{\ell} \e^2 \log(z \zb) {\tilde g}^{(4-\e)}_{\De_{\ell},\ell} (z,\zb)
+ \text{continuous at } z<0\,.
\end{align}

The discontinuity of $F(z, \zb)$ at $O(\e^2)$ reads:

\begin{align}\label{discf2}
\underset{z\, <\, 0}{\Disc}\, F^{(2)} (z,\zb) ={}& \pi i 
\left( C^{(1)}_{0} \gamma^{(1)}_{0} +  C^{MF}_{0} \gamma^{(2)}_{0} + \tfrac{1}{2}  C^{MF}_{0} (\gamma^{(1)}_{0})^2 \log(-z \zb) +  C^{MF}_{0} \gamma^{(1)}_{0} \partial_\e \right) {\tilde g}^{(4)}_{2,0} (z,\zb) \nonumber\\
& + 2 \pi i \sum\limits_{\substack{\ell=2\\\text{even}}}^\infty \frac{\Gamma(\ell+1)^2}{\Gamma(2\ell+1)}
\gamma^{(2)}_{\ell} {\tilde g}^{(4)}_{2+\ell,\ell} (z,\zb)\,.
\end{align}


We can evaluate the $\ell$ sum above using the explicit form for the conformal blocks in four dimensions:
\begin{align}
{\tilde g}^{(4)}_{2+\ell,\ell} (z,\zb) = \frac{k_{\ell+1}(z) - k_{\ell+1}(\zb)}{z-\zb}\
\end{align}
where $k_{\beta }(z)$ is defined in \eqref{block}. This results in:
\begin{align}
\sum\limits_{\substack{\ell=2\\\text{even}}}^\infty \frac{\Gamma(\ell+1)^2}{\Gamma(2\ell+1)\ell(\ell+1)}
{\tilde g}^{(4)}_{2+\ell,\ell} (z,\zb)&=\frac{1}{z-\zb}\bigg(\log(1-z) +\frac{1}{4} \log(1-z)^2 + \text{Li}_2(z)\nn&
-\log(1-\zb)- \frac{1}{4} \log(1-\zb)^2-\text{Li}_2(\zb)\bigg)\,.
\end{align}

Then, we evaluate the first order expansion of the conformal block using the expression for general dimension:

\begin{align}
{}&\partial_\e {\tilde g}^{(4)}_{2,0} (z,\zb) =
\frac{1}{z-\zb} \left(\frac{2}{3} \left(\text{Li}_2(z) -\text{Li}_2(\zb)\right) 
+\frac{1}{2} \left( \text{Li}_2 \hspace{-2pt} \left( \tfrac{\zb}{z} \right) -\text{Li}_2\hspace{-2pt} \left( \tfrac{z}{\zb} \right) + \text{Li}_2\hspace{-2pt} \left( \tfrac{z(1-\zb)}{\zb(1-z)} \right) -\text{Li}_2\hspace{-2pt} \left( \tfrac{\zb(1-z)}{z(1-\zb)} \right)
 \right)
\right.\nonumber\\
&+\left.
\frac{1}{2}\log\left(\tfrac{1-z}{1-\zb} \right)
\left(
\frac{4}{3} - \log(z-\zb)- \log(\zb-z)+ \log (z \zb) +\frac{1}{2} \log ((1-z)(1-\zb))
\right)
\right)\,.
\end{align}

Putting all the terms together in \eqref{discf2}, we finally compute the correlator from \eqref{dispfinal}:
\begin{align}
&F^{(2)} (z,\zb) = \frac{1}{z-\zb} \bigg[
-\frac{1}{12} \text{log} \left(\tfrac{z}{\zb}\right) \text{log} ^2\left(\tfrac{1-z}{1-\zb}\right)
-\frac{1}{12} \text{log} ^2((1-z) (1-\zb)) \text{log} \left(\tfrac{z}{\zb}\right)
\nonumber\\
&+\text{log} \left( \tfrac{1-z}{1-\zb} \right) \left(
\frac{10}{81} \text{log} \left( z \zb \right) 
+\frac{1}{12} \text{log}^2\left(\tfrac{z}{\zb}\right) 
-\frac{1}{36} \text{log} ^2(z \zb) 
-\frac{1}{9} \text{log} ((1-z) (1-\zb)) \text{log} (z \zb)
\right)
\nonumber\\
&-\frac{1}{18}
   (\text{Li}_2(z)-\text{Li}_2(\zb)) \left(4 \text{ log} ((1-z) z)+4 \text{ log} ((1-\zb) \zb)-\frac{40}{9}\right)
\label{eq:F2}\\
&+\frac{1}{3} \left( \left(\text{Li}_2\left(\tfrac{\zb-z}{\zb-1}\right) +\frac{1}{4} \text{log} ^2\left(\tfrac{1-z}{1-\zb}\right)\right) \text{log} (z
   \zb)
-\left(\text{Li}_2\left(\tfrac{\zb-z}{\zb}\right) +\frac{1}{4} \text{log}^2\left(\frac{z}{\zb}\right)\right)\text{log} ((z-1) (\zb-1))\right)
\nonumber\\
&+\frac{1}{3} \left(
\text{Li}_3\left(\tfrac{\zb-z}{\zb}\right)
-\text{Li}_3\left(\tfrac{z-\zb}{z}\right)
+\text{Li}_3\left(\tfrac{z-\zb}{z-1}\right)
-\text{Li}_3\left(\tfrac{\zb-z}{\zb-1}\right)
+\text{Li}_3\left(\tfrac{z-\zb}{z(1-\zb)}\right)
-\text{Li}_3\left(\tfrac{\zb-z}{\zb(1-z)}\right)\right) \bigg]\,.
\nonumber
\end{align}

\begin{figure}[h]

\begin{minipage}[b]{0.48\textwidth}
\begin{tikzpicture}[scale=0.6]
\draw   (-2,1) node[left ]{$x_1$}-- (-1,0)  ;
\draw  (-2,-1)node[left ] {$x_2$} -- (-1,0)  ;
\draw   (2,1) node[right]{$x_3$}-- (1,0)  ;
\draw   (2,-1) node[right]{$x_4$}-- (1,0)  ;
\draw (0,0) ellipse (1cm and 0.5cm);
\filldraw[black] (-1,0) circle (1.5pt) ;
\filldraw[black] (1,0) circle (1.5pt) ;
\filldraw[black] (-2,1) circle (1.5pt) ;
\filldraw[black] (-2,-1) circle (1.5pt) ;
\filldraw[black] (2,1) circle (1.5pt) ;
\filldraw[black] (2,-1) circle (1.5pt) ;
\put(80,2){$+\,\, {\rm permutations} \,\,\,\,+ $};
\end{tikzpicture}
\end{minipage}
\hfill
 \begin{minipage}[b]{0.48\textwidth}
\begin{tikzpicture}[scale=0.6]
\draw (10,0) circle (0.8cm);
\filldraw[black] (10,-0.8) circle (1.5pt) ;
\draw   (10,-0.8) -- (12,-0.8) node[right]{$x_4$ +\quad{\rm{permutations}}} ;
\filldraw[black] (12,-0.8) circle (1.5pt) ;
\draw   (10,-0.8) -- (8,-0.8)  ;
\filldraw[black] (8,-0.8) circle (1.5pt) ;
\draw   (8,-0.8) -- (6,-0.8) node[left]{$x_2$} ;
\filldraw[black] (6,-0.8) circle (1.5pt) ;
\draw   (8,-0.8) -- (6,-1.8) node[left]{$x_3$} ;
\filldraw[black] (6,-1.8) circle (1.5pt) ;
\draw   (8,-0.8) -- (6,0.2)  node[left]{$x_1$} ;
\filldraw[black] (6,0.2) circle (1.5pt)  ;
\end{tikzpicture}
\end{minipage}
\caption{Diagrams at $O(\e^2)$.}\label{fig:e2}
\end{figure}
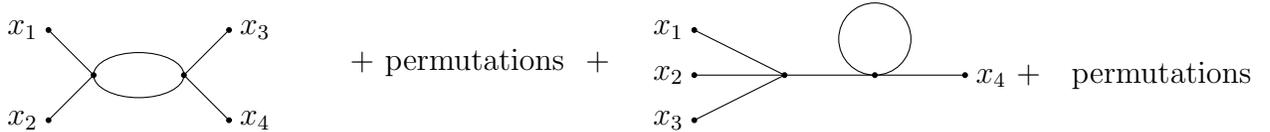

This correlator corresponds to the diagrams in Figure~\ref{fig:e2}.

Now, we decompose the correlator into conformal blocks to evaluate the OPE coefficients which is in agreement with~\cite{Gopakumar:2016cpb}:
\beq
C^{(2)}_{\ell} = \frac{(1+(-1)^\ell) \Gamma(\ell+1)^2}{\Gamma(2\ell+1)}
\frac{\ell (\ell+1) (H_{2\ell} - H_{\ell-1})-1}{9 \ell^2 (\ell+1)^2}\,,
\eeq
and~\cite{Alday:2017zzv}:
\beq
C^{MF}_{n,\ell} + \e^2 C^{(2)}_{n,\ell} = 
\begin{cases}
\frac{(1+(-1)^\ell)\Gamma(\ell+2)^2}{\Gamma(2\ell+3)}
\frac{\ell^2+3\ell+8}{24(\ell+1)(\ell+2)}
\left(
\frac{\e}{3}
\right)^2 + O(\e^3)\,, \quad &n=1\,,\\
O(\e^4)\,, \quad &n>1\,.
\end{cases}
\eeq

Here, $H_{\ell}$ is the harmonic number of order $\ell$.

To summarize, in~this section, we show that it is possible to use a dispersion relation which, together with crossing symmetry, specifies the OPE coefficients as a function of the conformal dimension, in~theories that admit a perturbative~expansion. 
\section{Basics of Superconformal Field~Theory}
\label{sec:SCFT}

We will now see what changes in this description when a theory is not only conformal invariant but also supersymmetric and we enter the realm of  superconformal field theories (SCFTs). The~presence of additional symmetry will even further constrain the spectrum of these theories and it will further help the analysis of correlation functions, unveiling new and interesting properties.  From~this moment onwards, we will focus on $d=4$ space--time dimensions, which will be relevant for the following\footnote{It is possible to prove that SCFTs can exist only for $d\leq 6$, for~$d\geq$ 7 is indeed not  possible to construct any Lie superalgebras satisfying certain consistency conditions~\cite{Nahm:1977tg,Minwalla:1997ka}.}.

When supersymmetry (SUSY) is in play, one needs to change and generalize the commutation relations in~\eqref{algebra} to allow for the presence of  supersymmetry generators, namely the supercharges $Q^{i}_{\alpha}$ and $\bar{Q}_{i \, \dot{\alpha}}$, with~$\alpha, \dot{\alpha}=1,2$ being the spinor indices. The~index $i$ takes values from $1$ to $ \mathcal{N}$, which corresponds to the  amount of supersymmetry; in four space--time dimensions and restricting to theories whose particles have spin up to $1$,  $\mathcal{N}$ ranges from 1 to 4, where the latter corresponds to  maximal supersymmetry.  By~studying the interplay between these supercharges and  the conformal generators, one realises that in order to have a  closed algebra, it is necessary to add some other fermionic generators, the~conformal supercharges $S\indices{_i^\alpha}$ and $\bar{S}\indices{^i^{\dot{\alpha}}}$.

Supersymmetry usually comes together with an additional symmetry that allows to rotate between supercharges, which is known as R-symmetry. Its generators $R^{i}_{j}$ organise in a $\mathfrak{u} (\mathcal{N})$ algebra for $\mathcal{N}=1,2,3$ or a $\mathfrak{su}(4)$ algebra  for $\mathcal{N}=4$.  A~distinguishing characteristic of SCFTs is that in this case, the R-symmetry it is not  an outer automorphism of the Poincar\'e SUSY algebra, as~it happens in the non-conformal supersymmetric case, but~it is really part of the algebra, as it commutes with the conformal subalgebra and acts non-trivially on the supercharges.  All in all, the~combination of conformal generators, $Q$'s, $S$'s and $R^i_j$ defines a simple Lie superalgebra:  $\mathfrak{su}(2,2|\mathcal{N})$ for $\mathcal{N}=1,2,3$ and $\mathfrak{psu}(2,2|4)$ for $\mathcal{N}=4$.  They contain, respectively, as bosonic subalgebras  $\mathfrak{so}(4,2)\times\mathfrak{su}(\mathcal{N})_R\times\mathfrak{u}(1)_R$ and $\mathfrak{so}(4,2)\times\mathfrak{su}(4)_R$, where in the first term of both expressions, we recognize the usual conformal algebra in four dimensions.  We will label operators based on how they transform under this  subset: we will specify  their conformal dimension $\Delta$, Lorentz quantum numbers $(j, \bar{j})$ and R-symmetry charges, encoded in the Dynkin labels of $\mathfrak{su}(\mathcal{N})_R$ and $\mathfrak{u}(1)_R$ charge when~present. 

Operators can be further organised into \textit{{superconformal primaries} 
} (or \textit{{superprimaries}}) and \textit{{superdescendants}}.  An~operator $\mathcal{O}$ is a  \textit{superprimary} if:
\begin{align}
S\indices{_i^\alpha}\ket{\mathcal{O}}=0 \, , \qquad\qquad \qquad \bar{S}\indices{^i ^{\dot{\alpha}}}\ket{\mathcal{O}}=0 \, .
\end{align}

Since schematically  $\{S,S\} \sim K$, a~superprimary is in particular a conformal primary, but~we notice that the converse is not true.  This implies that in a given superconformal representation, we can potentially find  many conformal primaries.  \textit{Superdescendants} are obtained  from a superprimary by acting with the $Q$'s, also in this case, the usual notion of conformal descendant is recovered thanks to the relation $\{Q,Q\}\sim P$. The~combination of a superprimary and all its descendants forms a superconformal~multiplet.

The last thing left to discuss is how the unitarity bounds in \eqref{ubound} change in the presence of SUSY. It can be argued that the presence of additional symmetry should determine even stronger~constraints. 

Suppose that there exists an unitarity bound given by some function $f(j, \bar{j}, R)$, where $R$ stands for $R$-symmetry quantum numbers, then we can distinguish~\cite{Cordova:2016emh,Minwalla:1997ka}:
\begin{itemize}
\item Superprimaries with $\Delta_{\mathcal{O}}> f(j, \bar{j}, R)$ which give rise to \textit{long multiplets}. In~general, a long multiplet contains $2^{4\mathcal{N}}$ states;
\item Operators at the unitarity bound and operators with $\Delta_{\mathcal{O}}< f(j, \bar{j}, R)$, but still allowed for specific spins and R-charges. These form \textit{short multiplets}, so called because they obey some ``shortening conditions'' that are concretely realized in the fact that they are annihilated by a certain amount of $Q$'s and $\bar{Q}$'s, and hence, the multiplet can only contain a reduced number of states.  These operators are often called \textit{BPS} and their dimension, being determined by Lorentz and R-symmetry quantum numbers, is protected against quantum corrections\footnote{The precise relation can be inferred by simple reasoning. Let us assume that the operator we want to consider is a superprimary, then in particular it holds $S \ket{\mathcal{O}}=0$. In~addition, it has to be annihilated by at least one supercharge, namely $Q \ket{\mathcal{O}}=0$.  This implies: 
\begin{align*}
0=\{S,Q\}\ket{\mathcal{O}}=(M^{\mu \nu}+D+R\indices{^i_j})\ket{\mathcal{O}}\sim (\mathcal{M}^{\mu \nu}+\Delta+R)\ket{\mathcal{O}}\, ,
\end{align*}
where $\mathcal{M}^{\mu \nu}$ encodes the Lorentz quantum numbers.
}.
\end{itemize}

\section{$\mathcal{N}$= 4 Super Yang--Mills}
\label{sec:N4}
In this section, we will analyse how the bootstrap techniques have been employed to study and powerfully constrain the spectrum and  correlators of $\mathcal{N}=4$ super Yang--Mills (SYM) theory  in four dimensions. One reason to study such theories is that, through the AdS/CFT correspondence~\cite{Maldacena:1997re, Gubser:1998bc,Witten:1998qj,Aharony:1999ti}, it is holographically related to Type IIB string theory on AdS$_5\times$S$^5$, where the CFT should be thought as living on the AdS boundary.   The~{precise form of the duality reads }

\begin{center}
    \begin{minipage}{.35\textwidth}
        \centering
\textbf{4d $\mathcal{N}=4$ SYM} with~$SU(N)$ gauge group and Yang--Mills coupling constant  $g_{ \text{YM}}$
    \end{minipage}%
    \begin{minipage}{.2\textwidth}
        \centering
\begin{tikzpicture}
\draw [{Latex[round]}-{Latex[round]}, thick] (-0.75,0) -- (0.75,0);
\end{tikzpicture}
    \end{minipage}%
    \begin{minipage}{.35\textwidth}
   \centering
\textbf{10d Type IIB string theory} on AdS$_5\times$ S$^5$ with string length $\sqrt{\alpha^{\prime}}$, $g_s$ coupling and radius $L=L_{\text{AdS}_5}=L_{\text{S}^5}$
    \end{minipage} \\ $ $

  \hspace{-0.43cm}
    $g_{ \text{YM}}^2 \qquad \quad =  \qquad \quad g_s$

  \hspace{-0.52cm}    $g_{ \text{YM}}^2 N \qquad \quad =  \qquad \quad\frac{L^4}{\alpha^{\prime}\,{}^2}$
 \end{center}

One of the checks that have been made to justify and investigate the validity of this  correspondence is verifying that the symmetries present on the two sides actually match. Furthermore, indeed, restricting to the bosonic part of the superconformal group,  we find that the $SO(2,4)$ conformal symmetry is reproduced by the isometries of AdS$_5$, while the R-symmetry group $SO(6)_R \simeq SU(4)_R$ is recovered as  the  isometries of the five-sphere. 

The AdS/CFT correspondence is conjectured to hold for any value of the parameters characterising the two theories and listed beforehand. However, it is useful to study two particular limits~\cite{DHoker:2002nbb}, which are the ones most widely used; for this purpose,  let us  introduce the \textit{'t~Hooft coupling} $\lambda \equiv g_{\text{YM}}^2 N$ \cite{tHooft:1973alw}.  First of all, let us take $N$ to infinity while keeping $\lambda$ fixed; in this limit, usually called \textit{'t Hooft limit},  the~large $N$ limit of the SCFT is mapped to weak coupling string perturbation theory, where each correction in powers of  $N^{-2}$ should be interpreted as a specific genus in the corresponding $g_s$ expansion.  Then, we can further take $\lambda \to \infty$, but~still less than $N$: in this second regime, the correspondence reduces to the one between strongly coupled  $\mathcal{N}=4$ SYM and Type IIB supergravity on weakly curved AdS$_5\times$S$^5$.  

Now, that we introduced the general framework in which $\mathcal{N}=4$ SYM can be inserted, let us delve further into the spectrum and properties of this~theory.

\subsection{Operators and~Spectrum}
$\mathcal{N}=4$ SYM  is believed to be, even if it is not yet proved, the~unique theory with the maximal possible amount of supersymmetry\footnote{If we restrict to quantum field theories containing at most spin 1 particles.} in four dimensions.  The~massless elementary fields of the theory are a gauge vector $A_{\mu}$, four Weyl fermions $\lambda_{\alpha}^a$ $(a=1 \ldots 4\, ,  \, \alpha=1,2)$ and six real scalars $\phi^i$ $(i=1 \ldots 6)$. They can all be rearranged to form a supermultiplet,  the~\textit{gauge multiplet}, and~as the name suggests, they transform into the adjoint of the $SU(N)$ gauge group.  Under the~R-symmetry group, $A_\mu$ is a singlet, $\lambda^a_{\alpha}$ transforms into the $\mathbf{4}$ of $SU(4)_R$ and $\phi^i$ into the fundamental of $SO(6)_R$ or equivalently as a rank 2 antisymmetric tensor of $SU(4)_R$.

Given the fundamental constituents of the theory, it is possible to write explicitly a Lagrangian~\cite{DHoker:2002nbb,Grimm:1977xp} and verify that this is indeed a conformal invariant and supersymmetric, at least classically. Quite remarkably, $\mathcal{N}=4$ SYM does not suffer from any perturbative UV divergences\footnote{Instantons corrections are believed to be UV finite as well.} at the loop level,  and as~a consequence, there is no need to introduce any scale during the renormalisation procedure, and hence, the $\beta$ function vanishes identically in the full quantum theory. This tells us that $\mathcal{N}=4$ SYM is exactly a superconformal field theory and $PSU(2,2|4)$ is a full quantum~symmetry. 

To classify the spectrum of the theory, we should construct all possible local, gauge invariant operators made of the canonical fields introduced above.  Among~these are superprimary operators, as defined in Section~\ref{sec:SCFT},  which can be constructed as the symmetric product of the elementary scalars $\phi^i$.  The~simplest configuration leads to \textit{single trace} operators of the form:
\begin{align}\label{singletrace}
\text{str}(\phi^{i_1} \cdots \phi^{i_n}) \, ,
\end{align}
where we are taking the symmetrized trace (str) over $SU(N)$, which makes the operator symmetric under the $SO(6)_R$ indices as well. In~general, \eqref{singletrace} defines a reducible representation and one has  to further distinguish between the trace and the traceless part of it.  In~the easiest example, by~doing so, we can differentiate between the \textit{Konishi operator}, $\sum_i\text{tr}(\phi^i \phi_i)$, and~$\mathcal{O}_2=\text{tr}\phi^{\{i}\phi^{j\}}$, where $\{ij\}$ singles out the traceless part. The products of these single trace operators constitute \textit{multi-trace} operators. 

As anticipated in Section~\ref{sec:SCFT}, it is convenient to classify states/operators according to the unitary representations of the bosonic subalgebra:
\begin{align}
\underbrace{\mathfrak{so}(1,3)}_{(j, \bar{j})}\quad \times\quad \underbrace{\mathfrak{so}(1,1)}_{\Delta}\quad\times\quad \underbrace{\mathfrak{su}(4)_R}_{[q, p, \bar{q}]} \, .
\end{align}

In addition to these quantum numbers, we will specify whether they satisfy some shortening conditions.  Interesting types of operators that will be relevant  for what follows~are: 
\begin{itemize}
\item \textit{Identity} operator, which is a singlet of R symmetry and it has $\Delta=0=\ell$;
\item \textit{$\frac{1}{2}$-BPS} operators, scalars annihilated by half of the supercharges. They can either be single trace operators:
\begin{align*}
\mathcal{O}_p(x)=\text{str} (\phi^{\{i_1}(x) \cdots \phi^{i_p\}}(x)) \qquad p\geq 2 \, ,
\end{align*}
symmetric traceless tensors transforming in the $[0,p, 0]$ or multi-trace operators:
\begin{align*}
\mathcal{O}_{(p_1 \cdots p_n )}(x)=\left[ \mathcal{O}_{p_1}(x) \cdots  \mathcal{O}_{p_n}(x) \right]_{[0,p,0]} \qquad \sum p_i = p \, ,
\end{align*}
where $[\, ]_{[q, p, \bar{q}]}$ stands for projection into the corresponding $SU(4)_R$ representation.
Their dimension $\Delta=p$ is protected from quantum corrections and supersymmetry completely fixes their three-point functions~\cite{Belitsky:2003sh,DHoker:2002nbb,Howe:1998zi,Dolan:2002zh,Freedman:1998tz,Lee:1998bxa,DHoker:1998vkc}.
\item \textit{$\frac{1}{4}$-BPS} operators $\left[ \mathcal{O}_{p_1}(x) \cdots  \mathcal{O}_{p_n}(x) \right]_{[q,p,q]} $,  with~Dynkin labels $[p,q,p]$ and protected dimension $\Delta=2p+q$ and  \textit{$\frac{1}{8}$-BPS} operators, multi-trace operators in the $[q,p, q+2m]$ having fixed dimension $\Delta=p+2q+2m$. Both these types are genuinely BPS only in the free theory and mix with descendants of non-BPS operators when interactions are turned on. 
\item \textit{Long} operators can transform into a generic $[q, p, \bar{q}]$ R-symmetry representation, as their dimension is not protected but nonetheless subject to the unitarity bound:
\begin{align}
\Delta \geq \text{max}\left(2+2j+\frac{3}{2}q+p+\frac{\bar{q}}{2},2+2\bar{j}+\frac{3}{2}\bar{q}+p+\frac{q}{2}\right) \, .
\end{align}
\end{itemize} 

In light of the AdS/CFT correspondence, it is possible to establish a dictionary between these operators and fields in the dual AdS, after~having compactified along S$^5$  \cite{Andrianopoli:1998jh, Maldacena:1997re}. Single trace operators $\mathcal{O}_p$ are mapped to single particle states,  in~particular $\mathcal{O}_2$, which is dual to the scalar of the graviton supergravity multiplet, while $\mathcal{O}_{p\geq 3}$ correspond to its Kaluza Klein modes. The~masses of these supergravity scalars are completely fixed by the conformal dimension of their duals as $m^2=\Delta(\Delta-4)$.  Multi-trace operators, on~the other side, are mapped to threshold multiparticle bound states in AdS. Let us conclude by mentioning that it can be shown that some other non-BPS operators, such as the Konishi multiplet,  scale as $\lambda^{1/4}$: these operators correspond to massive string modes that decouple in the supergravity regime $\lambda \to \infty$, which is the~one we are mainly interested~in.

\subsection{Stress Tensor Multiplet~Correlators}

As we were mentioning before, the two- and three-point functions of $\frac{1}{2}$-BPS operators are completely fixed by superconfomal symmetry and do not receive quantum corrections thanks to some {non renormalisability} theorems~\cite{Intriligator:1998ig,Intriligator:1999ff,Eden:1999gh,Petkou:1999fv,Howe:1999hz,Heslop:2001gp}. The~four-point function instead enjoys only partial renormalisability and therefore it does depend on the coupling constant  but just through a trivial function~\cite{Bissi:2015qoa,Korchemsky:2015ssa,Belitsky:2014zha,Beem:2013qxa,Beem:2016wfs}. Thanks to this property, the~four-point function of any superdescendant is determined by the one of the corresponding superprimary. Therefore, it is enough to study  the correlators of  superprimaries, in~general, easier since they involve  scalars, to~constrain correlators of superdescendants.  This turns out to be incredibly useful if one wants to study supergravity.  Studying graviton amplitude amounts to compute correlators of the stress tensor is generally difficult to do. However, the stress tensor belongs, together with R-symmetry and super currents, to~a $\frac{1}{2}$-BPS multiplet, whose superprimary is the $\mathcal{O}_2$ single trace operator introduced before.  Thus,  we can focus on $\langle \mathcal{O}_2(x_1)\mathcal{O}_2(x_2)\mathcal{O}_2(x_3)\mathcal{O}_2(x_4)\rangle$ and that is why we will devote the rest of the section to its analysis.  In~particular, we will see how to ``bootstrap'' this correlator, meaning employing all the available symmetries to constrain its form and obtaining information about the spectrum of our theory~\cite{Dolan:2001tt,Nirschl:2004pa,Dolan:2004iy, Beem:2013qxa,Beem:2016wfs}.

$\mathcal{O}_2$ is a scalar of protected dimension $\Delta=2$ and it transforms into the $[0,2,0]=\mathbf{20^\prime}$ {representation} of an R-symmetry group, thus as a symmetric traceless tensor. To~ensure this one can introduce $SO(6)$, null vectors $t^i$, $i=1\ldots 6$, $t \cdot t=0$ and rewrite:
\begin{align}
\mathcal{O}_2(x, t)=t^i t^j \text{tr} (\phi^i(x) \phi^j(x)) \, .
\end{align}

As already mentioned, it is the superconformal primary of the supermultiplet to which the stress energy tensor~belongs. 

Conformal symmetry on its own already partially fixes the form of  its four-point function to be: 

\begin{align}
\langle \mathcal{O}_2(x_1, t_1)\mathcal{O}_2(x_2, t_2)\mathcal{O}_2(x_3, t_3)\mathcal{O}_2(x_4, t_4)\rangle=\left( \frac{t_1 \cdot t_2 \, t_3 \cdot t_4}{x_{12}^2 x_{34}^2} \right) ^2 \mathcal{F}({t_i}, u, v)\, ,
\end{align}
where $u$ and $v$ are the conformal cross ratios introduced in \eqref{crossrat}. Then, we can enforce $SU(4)_R$ symmetry: this imposes constraints on the possible representations that can be exchanged in  the OPE:
\begin{align}\label{exchangedRepr}
[0,2,0]\otimes [0,2,0]=\underbrace{[0,0,0]}_{\mathbf{1}} \oplus\underbrace{[1,0,1]}_{ \mathbf{15}}\oplus\underbrace{[0,2,0]}_{ \mathbf{20^\prime}}\oplus\underbrace{[2,0,2]}_{ \mathbf{84}}\oplus \underbrace{[0,4,0]}_{\mathbf{105}}\oplus \underbrace{[1,2,1]}_{\mathbf{175}} \, .
\end{align}

Such decomposition allows us to rewrite:

\begin{align} \begin{aligned}
\mathcal{F}(t_i, u, v)&=\sum_{0 \leq m \leq n\leq 2} A_{nm}(u,v) Y_{nm} (\sigma, \tau)\,  ,\\ Y_{nm} (\sigma, \tau)&=\frac{P_{n+1}(y)P_m(\bar{y})-P_{m}(y)P_{n+1}(\bar{y})}{y-\bar{y}}\, ,
\end{aligned}
\end{align}
where $n,\, m$ label the six representations $[n-m, 2m, n-m]$ exchanged  in  \eqref{exchangedRepr}.  The~functions  $Y_{nm}$ are  $SO(6)_R$ harmonics, which can be written in terms of Legendre polynomials $P_n$
and depend on the polarization cross ratios~\cite{Dolan:2004iy}:
\begin{align}
\sigma&= \frac{t_1 \cdot t_3 \, t_2 \cdot t_4}{t_1 \cdot t_3 \, t_2 \cdot t_4}=\alpha \bar{\alpha}=\frac{(1+y)(1+\bar{y})}{4}\, , \\ 
 \tau&= \frac{t_1 \cdot t_4 \, t_2 \cdot t_3}{t_1 \cdot t_3 \, t_2 \cdot t_4}=(1-\alpha)(1- \bar{\alpha})=\frac{(1-y)(1-\bar{y})}{4}\, .
\end{align}

Notice that for the correlator of four, dimension two, operators $\mathcal{F}$ is a polynomial of degree 2 in $\sigma$ and $\tau$.

Since we are considering the correlation function of identical operators, we need to impose invariance under permutations of all external operators. This translates into the following crossing equations for the function $\mathcal{F}$:
\begin{align}\label{crossingF}
\begin{aligned}
\mathcal{F}(u,v,\sigma, \tau)&=\left(\frac{u}{v}\right)^2 \tau^2 \mathcal{F}\left(v,u,\frac{\sigma}{\tau}, \frac{1}{\tau}\right)&& \qquad 1 \leftrightarrow 3 \text{ exchange}\\
&= \mathcal{F}\left(\frac{u}{v},\frac{1}{v},\tau, \sigma \right)&& \qquad 1 \leftrightarrow 2 \text{ exchange}
\end{aligned}
\end{align}

These requirements can be read as well as consistency conditions for the $A_{nm}(u,v)$, which admits an expansion in the usual conformal blocks\footnote{In the literature, this expansion is also called conformal partial wave expansion or conformal partial wave amplitude.}:
\begin{align}
A_{nm}(u,v)= \sum_{\Delta,\ell} A_{nm\, , \, \ell}^{\Delta} u^{\frac{\Delta-\ell}{2}}\tilde{g}_{\Delta, \ell}(u,v)\,  , 
\end{align}
 where the 4D conformal blocks have been defined in \eqref{cb} and \eqref{gd}\footnote{With respect to these expressions, we suppressed the superscript $(d=4)$ in the definition of the blocks since it is assumed that we are working in four dimensions.}.  Notice that the sum runs over the spin $\ell$ because in the OPE of two  scalars, the~exchanged operators can only be symmetric traceless Lorentz tensors, for~which $j=\bar{j}=\frac{\ell}{2}$ with $\ell$ even.  As~already discussed before expanding, conformal blocks allow us to packing together the contribution of each conformal primary and all its descendants. However, in the presence of supersymmetry\footnote{All operators in a superconformal multiplet must have the same anomalous dimension.}, we would like to expand the correlator in such a way that each supermultiplet contributes to the OPE as a whole; in~other words, we would like to find some  \textit{superconformal blocks} condensating the contribution of each superprimary and all its descendants. This is in general very hard, however, it is possible in this case by  fully exploiting the power of superconformal invariance and solving \textit{superconformal Ward identities}.  These identities can be phrased as~\cite{Dolan:2004iy}
\begin{align}\label{SWI}
\mathcal{F}(z, \zb, \alpha, \bar{\alpha})\Big|_{\bar{\alpha}=\frac{1}{\zb}}=f(z, \alpha)\, ,
\end{align}
with analogous requirements  for $\alpha \leftrightarrow \bar{\alpha},\,  z\leftrightarrow \zb$. The~function $f$ encodes the contribution from the protected sector of the theory, namely from all the possible short and semi-short representations that can be exchanged in the OPE,  and~for this reason, can be completely determined by free field theory results. It has to satisfy the consistency condition $f(z, 1/z)=k$, where $k$ is a generic constant and in light of this, it can be rewritten as
\begin{align}
f(z, \alpha)=k+\left(\alpha+\frac{1}{z} \right)\hat{f}(z, \alpha) \, .
\end{align}

In a complementary interpretation~\cite{Beem:2013sza,Beem:2016wfs}, $f(z, \alpha)$ arises from the appearance of an additional chiral structure, which is a general property of $\mathcal{N}=2$ superconformal field theories but can be extended to $\mathcal{N}=4$ as well.  In~this picture, $f(z, \alpha)$ can be understood as a correlator of a two-dimensional auxiliary chiral~algebra. 

A generic solution of the Ward identity in \eqref{SWI} can be written as

\begin{align}
&\mathcal{F}(u,v,\sigma, \tau)=\mathcal{F}|_{\hat{f}}(z, \zb, \alpha, \bar{\alpha})+ (1-z \alpha)(1-\zb \alpha)(1-z\bar{\alpha})(1-\zb\bar{\alpha})\mathcal{G}(z, \zb) \, ,\\ \nonumber
&\mathcal{F}|_{\hat{f}}=-k+\frac{(1-\zb \alpha)(1-z \bar{\alpha})\left[f(z, \alpha)+f(\zb, \bar{\alpha})\right]-(1-z \alpha)(1-\zb \bar{\alpha})\left[f(z, \bar{\alpha})+f(\zb, \alpha)\right]}{(z-\zb)(\alpha-\bar{\alpha})} \, .
\end{align}

$\mathcal{G}(z, \zb)$ contains the dynamical information of the theory and encodes the contribution from long supermultiplets. In~the special case of external dimension $p=2$, it does not depend on the $SU(4)_R$ cross ratios.  Remarkably, it is possible for it to find a  decomposition in superconformal blocks:
\begin{align}\label{expBlocks}
\mathcal{G}(u,v)=\sum_{\Delta, \ell} A_{\Delta,\ell} u^{\frac{\Delta-\ell}{2}}\tilde{g}_{\Delta+4, \ell}(u,v)\, ,
\end{align}
where it turns out that a superconformal block is just a usual block with a shift by 4 in the dimension. The~exchanged operators are long supermultiplets whose lowest dimension operator is a singlet of $SU(4)_R$. Furthermore, the function $\hat{f}(z, \alpha)$ admits a similar expansion:
\begin{align}
\hat{f}(z, \alpha)=\sum_{\ell=0}^{\infty}b_{0,\ell}\, \tilde{g}_{\ell+2}(z)P_0(y)+\sum_{\ell=-1}^{\infty}b_{1,\ell}\, \tilde{g}_{\ell+2}(z)P_1(y)\, ,
\end{align}
where the coefficients $b_{i, \ell+2}$ are known~\cite{Dolan:2004iy} and we introduced $\mathfrak{sl}(2)$ blocks $\tilde{g}_{\ell}(z)=\left(\frac{1}{2}z\right)^\ell {}_2F_1(\ell, \ell, 2\ell, z)$. Here, only short representations, whose dimension can be fixed in terms of the spin, contribute. These decompositions in superconformal blocks make manifest the contributions of the various types of multiplets. However, there is still an intrinsic ambiguity due to the fact that at the unitarity threshold, long multiplets decompose into short and semi-short ones and in a non-interacting theory, there is no way to distinguish truly protected from unprotected contributions. It is indeed common to further distinguish:
\begin{align}
\mathcal{G}(u,v)=\mathcal{G}^{short}(u,v)+\mathcal{H}(u,v)\,  ,
\end{align}
where we stripped out the contributions coming from the protected sector in $\mathcal{G}^{short}(u,v)$, which is explicitly known, as can be seen in~\cite{Beem:2016wfs}.  $\mathcal{H}(u,v)$ can still be expanded as in \eqref{expBlocks}:
\begin{align}\label{BlockExpH}
\mathcal{H}(u,v)= \sum_{\substack{\Delta, \ell \\ \ell \text{ even}}}a_{\Delta, \ell}u^{\frac{\Delta-\ell}{2}}\tilde{g}_{\Delta+4, \ell}(u,v)\,,
\end{align}
where now the sum is over unprotected long operators, singlet of R-symmetry, with~$\Delta \geq \ell+2$ and $a_{\Delta, \ell}\geq 0$, as expected from~unitarity. 

The crossing conditions in \eqref{crossingF} becomes:
\begin{align}\label{crossH}
v^2\mathcal{H}(u,v)-u^2 \mathcal{H}(v,u)=-v^2\mathcal{G}^{short}(u,v)+u^2 \mathcal{G}^{short}(v,u)-(u^2-v^2)-\frac{u-v}{c}\, ,
\end{align}
where $c=\frac{N^2-1}{4}$ is the central charge of the theory and $\mathcal{G}^{short}$ is linear in $1/c$.

To date, we have seen how symmetries give stringent constraints on the form of the four-point function and can already give information about the protected spectrum of the theory.  However, there are still undetermined data hidden  in $\mathcal{H}$ and one would like to find a way to study the dimensions and the squared OPE coefficients $a_{\Delta, \ell}$ appearing there.  Various approaches have been pursued in the years, from~numerical bootstrap   techniques \cite{Beem:2013qxa, Beem:2016wfs,Alday:2013opa, Alday:2014qfa,Bissi:2020jve} to the more analytic ones~\cite{Aprile:2017xsp,Aprile:2017bgs,Alday:2017xua,Alday:2017vkk,Aprile:2018efk,Alday:2018pdi,Alday:2018kkw,Rastelli:2016nze,Rastelli:2017udc,Alday:2019nin,Caron-Huot:2018kta,Bissi:2020wtv,Bissi:2020woe}, involving  the use of modern tools such as the Lorentzian inversion formula~\cite{Caron-Huot:2017vep,Simmons-Duffin:2017nub,Kravchuk:2018htv}, large spin perturbation theory~\cite{Alday:2016njk} and unitarity methods~\cite{Aharony:2016dwx,Meltzer:2019nbs,Antunes:2020pof}.  These studies have shed some light on the spectrum of long operators in $\mathcal{N}=4$ and they have provided non-trivial tests of the AdS/CFT correspondence, especially in the large $N$ (or equivalently large $c=\frac{N^2-1}{4}$) limit and at infinite 't Hooft coupling $\lambda$.  In~this limit, the~interacting part of the correlator can be expanded as
\begin{align}\label{largeNH}
\mathcal{H}(z, \zb)= \sum_{\kappa=0}^{\infty} \frac{\mathcal{H}^{(\kappa)}(z, \zb)}{c^{\kappa}}\, ,
\end{align}
where each term  maps to a $(\kappa-1)$ loop in the dual gravity amplitude\footnote{$\mathcal{H}^{(0)}$ maps to the disconnected part of the amplitude}.

As discussed before,  in~this regime,  corresponding to the supergravity approximation, all string modes become infinite massive and we are left with protected single trace operators and long multi-particle ones, which are dual to multi-trace operators in the dual picture. The~latter receive corrections both to their dimensions and OPE coefficients at  large $N$.  Among~them, the only ones with  a non-zero anomalous dimension and three-point coefficient already at order $c^{-1}$ are \textit{double trace} operators\footnote{The other multi-trace operators get corrections at order $c^{-2}$ and higher.}. They are constructed from the product of $\frac{1}{2}$-BPS operators and they take the schematic form  $[\mathcal{O}_p\, \mathcal{O}_p]_{n,\ell}=(\mathcal{O}_p \Box^{n}\partial_{\mu_1}\dots\partial_{\mu_\ell} \mathcal{O}_p -\text{traces})$ and at leading order, and they  assume their classical dimension is $\Delta= 2p+2n+\ell$.  Their OPE data admit an expansion similar to \eqref{largeNH}:
\begin{align}
\label{largeNexp}
\tau_{n, \ell}&=4+2n +\frac{1}{c} \gamma_{n,\ell}^{(1)}+\frac{1}{c^2} \gamma_{n,\ell}^{(2)}+ \dots \, ,\\ \label{largeNexp2}
a_{n,\ell}&=a_{n,\ell}^{(0)}+\frac{1}{c}a_{n,\ell}^{(1)}+\frac{1}{c^2}a_{n,\ell}^{(2)}+\dots \, ,
\end{align}
where we introduced the twist $\tau=\Delta-\ell$ and we denoted with $\gamma^{(\kappa)}$ the anomalous dimension at order $c^{-\kappa}$.   Notice that for fixed $n$ and $\ell$ there, can be more than one superconformal primary with the same twist and transforming in the same    $SU(4)_R$ representation. 

By plugging \eqref{largeNexp} and \eqref{largeNexp2} in the expression for $\mathcal{H}(u,v)$ in \eqref{BlockExpH}, we can express the single $\mathcal{H}^{(\kappa)}$ in terms of anomalous dimensions and corrections to the OPE coefficients. The~first few terms are given by
\begin{align}
\mathcal{H}^{(0)}&=\sum_{n, \ell}u^{n+2}a_{n, \ell}^{(0)}\tilde{g}_{2n+8, \ell}(z, \zb)\, , \\ \label{H1}
\mathcal{H}^{(1)}&=\sum_{n, \ell}u^{n+2}\left(a_{n, \ell}^{(0)} \gamma_{n, \ell}^{(1)}\partial_n+a_{n, \ell}^{(1)} +\frac{1}{2}\log u\, a_{n, \ell}^{(0)} \gamma_{n, \ell}^{(1)} \right) \tilde{g}_{2n+8, \ell}(z, \zb)\, , \\ \label{H2}
\mathcal{H}^{(2)}&=\sum_{n, \ell}u^{n+2}\left\lbrace  a_{n, \ell}^{(0)}\left(  \frac{1}{2}(\gamma_{n, \ell}^{(1)}   )^2 \partial_n^2 +\gamma_{n, \ell}^{(2)}\partial_n \right ) +a_{n, \ell}^{(1)}\gamma_{n, \ell}^{(1)}\partial_n+a_{n, \ell}^{(2)} \right. \\ & \nonumber \left. +\frac{1}{2}\log u \left[ a_{n, \ell}^{(0)}\gamma_{n, \ell}^{(2)}+a_{n, \ell}^{(1)}\gamma_{n, \ell}^{(1)}+a_{n, \ell}^{(0)}(\gamma_{n, \ell}^{(1)})^2\partial_n \right]   +\frac{1}{8} \log ^2 u \,a_{n, \ell}^{(0)}(\gamma_{n, \ell}^{(1)})^2\right\rbrace \tilde{g}_{2n+8, \ell}(z, \zb) \, .
\end{align}

The quantity in the first line can be derived from disconnected diagrams in free field theory and  allows to fix:
\begin{align}
\label{averageA0}
&a_{n,\ell}^{(0)}=\frac{\pi (\ell+1)(\ell+2n+6)\Gamma(n+3)\Gamma(\ell+n+4)}{2^{(2\ell+4n+9)}\Gamma(n+\frac{5}{2})\Gamma(\ell+n+\frac{7}{2})}\, .
\end{align} 

An explicit expression is also known at order $c^{-1}$ in terms of so-called $\overline{D}$   function \cite{Dolan:2000ut, DHoker:1999kzh}:
\begin{align}
\mathcal{H}^{(1)}(u,v)=-u^2 \overline{D}_{2422}(z, \zb)= h(u,v) \log u+ \tilde{h}(u,v)\, ,
\end{align}
where both $h$ and $\tilde{h}$ admit an expansion in power of $u$ if we  allow for negative powers  in $\tilde{h}(u,v)$. By~matching $h(u,v)$ with the logarithmic part in \eqref{H1}, one can extract the anomalous dimension of double trace operators at order $c^{-1}$:
\begin{align}\label{averageGamma}
\gamma_{n, \ell}^{(1)}&=-\frac{(n+1)(n+2)(n+3)(n+4)}{(\ell+1)(2n+\ell+6)}\, ,
\end{align}
and analogously, from $\tilde{h}$, one can obtain:
\begin{align}\label{averageA1}
a_{n, \ell}^{(1)}&=\frac{1}{2}\partial_n \left( a_{n, \ell}^{(0)} \gamma_{n, \ell}^{(1)}\right)\, .
\end{align}

Going to a higher order, we inevitably run into problems due to  mixing among degenerate  double trace operators, since  all operators of the form  $[\mathcal{O}_2\mathcal{O}_2 ]_{n,\ell}, [\mathcal{O}_3\mathcal{O}_3 ]_{n-1,\ell}, \dots, $ $[\mathcal{O}_{n+2}\mathcal{O}_{n+2} ]_{0,\ell}$  will equally contribute to \eqref{largeNexp}.  In~light of this,  the~tree level information only fixes for us the averages, over~all these degenerate states, of~the various OPE data; however, at one loop (and higher), powers and products of these data appear, so that computing them requires  unmixing the different contributions.  Therefore, let us introduce an additional index $I$ to account for the degeneracy and  define  new $\mathcal{O}_{n, \ell, I}$ with $I=1, \dots, n+1$ to be eigenfunctions of the dilatation operator.  Solving this mixing problem then requires computing $a^{(\kappa)}_{n, \ell, I}$ and $\gamma^{(\kappa)}_{n, \ell, I}$ for each index $I$, order-by-order in the large $c$ expansion. Remarkably, this has been done at order $c^0$ and partially at $c^{-1}$ by studying mixed correlators  $\langle\mathcal{O}_p\mathcal{O}_p\mathcal{O}_q\mathcal{O}_q\rangle$ in~\cite{Aprile:2017bgs,Aprile:2017xsp,Alday:2018pdi,Alday:2017xua}, providing as with explicit expressions for $a_{n, \ell, I}^{(0)}$ and $\gamma_{n, \ell, I}^{(1)}$.

A closer look to \eqref{H2} shows us that this knowledge  is sufficient to completely reconstruct one piece of the correlator, namely the leading $\log u$ one:
\begin{align}\label{H2HigherLog}
\mathcal{H}^{(2)}(u, v)\Big|_{\log^2 u}=\frac{1}{8}\sum_{n, \ell}\sum_{I=1}^{n+1}(z\zb)^{n+2}a_{n, \ell, I}^{(0)}(\gamma_{n, \ell, I}^{(1)})^2 \tilde{g}_{2n+8, \ell}(z, \zb)\, ,
\end{align}
where this expression can be explicitly resummed in terms of logarithms and polylogarithms times some rational functions~\cite{Aprile:2017bgs}.  However, the importance and relevance of this term is not limited to the possibility of  computing it, which is still quite extraordinary; rather, it relies on the  fact that this is enough to extract information on $\gamma^{(2)}$ and eventually reconstruct the full four-point function $\mathcal{H}^{(2)}$ \cite{Alday:2017xua,Alday:2017vkk,Aprile:2017bgs}. We will now try to briefly discuss how this is concretely~realised.  

First of all, it is possible  to show that in the small $v$ limit, \eqref{H2HigherLog} behaves as
\begin{align}\label{H2smallV}
\mathcal{H}^{(2)}(u,v)\Big|_{\log ^2 u} \sim p(u,v) \log^2 v +\tilde{p}(u,v) \log v +\text{regular terms}\, ,
\end{align}
where $p$ and $\tilde{p}$ are known polynomials.
Now remember that at order $c^{-2}$ and higher, the crossing symmetry condition \eqref{crossH} simply reads $v^2 \mathcal{H}^{(k\geq 2)}(u,v)=u^2 \mathcal{H}^{(k\geq 2)}(v,u)$. Putting these two together, we have that in the first term of \eqref{H2smallV}, which is crossing symmetric on its own,  $p(u,v)$ should satisfy:
\begin{align}
v^2 p(u,v)=u^2 p(v,u) \, ,
\end{align}
while the second term is schematically mapped to: 
\begin{center}
$v^2 \mathcal{H}^{(2)}(u,v)\Big|_{\log ^2 u\, \log v} \xleftrightarrow{\hspace{0.5cm}\text{crossing}\hspace{0.5cm}} u^2 \mathcal{H}^{(2)}(v,u)\Big|_{\log ^2 v\, \log u}$\\
\hspace{5cm}\rotatebox[origin=c]{90}{$\subset$} \\ \vspace{0.2cm}
\hspace{6cm}$\frac{1}{2}(a_{n, \ell}^{(0)}\gamma_{n, \ell}^{(2)}+a_{n, \ell}^{(1)}\gamma_{n, \ell}^{(1)}+a_{n, \ell}^{(0)}(\gamma_{n, \ell}^{(1)})^2\partial_n)\tilde{g}_{2n+8, \ell}$
\end{center}

This diagram tells us that through crossing,  $\tilde{p}(u,v)$ contains information about the unknown one loop anomalous dimensions and gives us a concrete procedure to extract~them.

The other interesting property of  \eqref{H2HigherLog} is that it is the only term with a non-vanishing double discontinuity appearing in the correlator at this order. The~double discontinuity, dDisc for short, is defined  as the difference between the Euclidean correlator and its two possible analytic continuations around $\zb=1$, keeping $z$ fixed:
\begin{align}
\label{dDiscLog}
\text{dDisc} \mathcal{H}(z, \zb) \equiv \mathcal{H}(z, \zb) -\frac{1}{2}\left( \mathcal{H}^{\circlearrowleft}(z,\zb)+\mathcal{H}^{\circlearrowright}(z, \zb) \right) \, .
\end{align}

Using this definition, it is not difficult to check that when applying dDisc to positive integer powers of  $(1-\zb)$, $\log(1-\zb)$ and  their product, we obtain  zero, hence the only piece surviving  in $\mathcal{H}^{(2)}$ is the one proportional to $\log^2(1-\zb)$. Through crossing, this corresponds to the $\log^2 u$ term we are considering and for which we know an explicit expression. The~reason why this quantity is so interesting and that the idea that it can be fixed by the leading log term in \eqref{H2HigherLog} is so appealing is that dDisc represents the only necessary ingredient of the Lorentzian inversion formula~\cite{Caron-Huot:2017vep}, which provides an alternative and parallel way to extract the full OPE data and to reconstruct entirely the~correlator.

At the same time, the discussion in terms of dDisc opens new ways of interpreting CFT correlation functions, especially in relationship with their dual gravity amplitudes.  It has been shown, and~explicitly checked at one loop~\cite{Alday:2017vkk}, that it is possible to relate the double discontinuity of $\mathcal{H}$ in a certain kinematic limit, called \textit{flat space limit} \cite{Alday:2018pdi, Heemskerk:2009pn,Okuda:2010ym, Maldacena:2015iua, Gary:2009ae,Susskind:1998vk,Polchinski:1999ry},  to~the discontinuity of the corresponding supergravity amplitude computed in $\mathbb{R}^ {10}$.  At~order $c^{-2}$, this can be nicely {summarised as} 
\begin{align}
a^{(0)}(\gamma^{(1)})^2 \sim \text{dDisc}\,\mathcal{H}^{(2)} \xLeftrightarrow[\text{limit}]{\text{flat space}}\begin{tikzpicture}[baseline={([yshift=-1.1ex]current bounding box.center)},scale=0.3,node/.style={draw,shape=circle,fill=black,scale=0.4}]
   \draw [thick] (0,0) -- (3,0)--(3,3)--(0,3)--(0,0);
    \draw [thick] (0,0) -- (-1,-1);
    \draw [thick] (0,3) -- (-1,4);
    \draw [thick] (3,0) -- (4,-1);
    \draw [thick] (3,3) -- (4,4); 
    \draw [dashed,thick,orange] (1.5,4)--(1.5,-1);
  \end{tikzpicture} \, .
\end{align}

On the RHS, we pictured the one-loop graviton amplitude as its Feynman diagram (sum over all other permutations is understood); the dashed vertical lines, cutting the diagram in two parts,  reflect the fact that we are taking a discontinuity~\cite{Cutkosky:1960sp}.

Establishing a connection between dDisc and discontinuities represents another important building block in the more general attempt to understand how unitary techniques, well known and established in the amplitude context, can be adapted and translated for CFT correlation functions~\cite{Aharony:2016dwx, Meltzer:2019nbs, Meltzer:2020qbr}. For~these purposes, it would be interesting to check this correspondence and generalise the previous discussion to higher  orders in the $1/c$ expansion.  This program has been initiated in~\cite{Bissi:2020wtv, Bissi:2020woe}, and here it has been shown that from two loops onward,  the~knowledge of $a^{(0)}_{n, \ell, I}$ and $\gamma^{(1)}_{n, \ell, I}$ is no longer sufficient to completely fix the correlator, but it is still interesting and can suggests new interplays between correlators and amplitude singularities. Let us focus on $\mathcal{H}^{(3)}$ for simplicity, as this has an expansion in conformal blocks analogous to \eqref{H1} and \eqref{H2}.  In~particular, as~it happens at one loop, there is a term depending only on tree level OPE data, namely:

\begin{align}\label{H3HigherLog}
\mathcal{H}^{(3)}(u, v)\Big|_{\log^3 u}=\frac{1}{48}\sum_{n, \ell}\sum_{I=1}^{n+1}(z \zb)^{n+2}a_{n, \ell, I}^{(0)}(\gamma_{n, \ell, I}^{(1)})^3 \tilde{g}_{2n+8, \ell}(z, \zb)\, ,
\end{align}
which can be resummed by giving an expression in terms of functions that are  generalizations of classical polylogarithms. In contrast to the case $c^{-2}$, this term does not saturate the full dDisc,  and $\mathcal{H}^{(3)}$ indeed contains a term proportional to $\log^2 u$, which depends on OPE data (such as $\gamma^{(2)}$) for which the mixing has not yet been solved.  This fact prevents us from being able to fully reconstruct the correlation function.  At~the same time, there is  another important  source of complication that comes from the appearance of higher trace operators  in the terms of the correlator with non-vanishing double discontinuity starting at order $c^{-3}$.  In~particular, at~two loops, triple trace operators start  mixing with the double trace ones we considered so far, so one should in principle find a way to treat them and to disentangle their contributions from the known ones in order to constrain the form of the four-point~function.

Nonetheless, the expression in \eqref{H3HigherLog}  finds a specific counterpart in the dual supergravity amplitude: it can be shown that the dDisc restricted to the leading log term can be mapped to a ``double-cut'' of the planar two-loop four graviton amplitude. Pictorially, \mbox{{this reads}}

\begin{align}
a^{(0)}(\gamma^{(1)})^3 \sim \text{dDisc}\,\mathcal{H}^{(3)}\Big|_{\log^3 u} \xLeftrightarrow[\text{limit}]{\text{flat space}}\begin{tikzpicture}[baseline={([yshift=-1.1ex]current bounding box.center)},scale=0.3,node/.style={draw,shape=circle,fill=black,scale=0.4}]
   \draw [thick] (0,0) -- (6,0)--(6,3)--(0,3)--(0,0);
    \draw [thick] (0,0) -- (-1,-1);
    \draw [thick] (0,3) -- (-1,4);
    \draw [thick] (6,0) -- (7,-1);
    \draw [thick] (6,3) -- (7,4); 
     \draw [thick] (3,0) -- (3,3); 
    \draw [dashed,thick,orange] (1.5,4)--(1.5,-1);
    \draw [dashed,thick,orange] (4.5,4)--(4.5,-1);
  \end{tikzpicture} \, .
\end{align}

Similar conclusions can be drawn to all order in the large $c$ expansion.

To conclude, let us  briefly mention related studies that can be found in the literature.  Similar analyses to the one presented above were performed in  Mellin space   in \cite{Fitzpatrick:2011ia,Rastelli:2016nze,Rastelli:2017udc,Alday:2018kkw,Alday:2019nin}.  More generic configurations of the correlator have been considered, for~instance \textls[-25]{by allowing for external operators with  different $p_i$ \cite{Arutyunov:2002fh,Arutyunov:2003ae,Berdichevsky:2007xd,Uruchurtu:2011wh,Aprile:2017xsp,Aprile:2018efk,Caron-Huot:2018kta}. Finally,  $\langle\mathcal{O}_2\mathcal{O}_2\mathcal{O}_2\mathcal{O}_2\rangle$} was investigated separately from the strict supergravity limit in a series of works~\cite{Alday:2018pdi, Drummond:2019hel,Drummond:2020dwr, Aprile:2020luw} where $\alpha^{\prime}$ or equivalently $\frac{1}{\sqrt{\lambda}}$ string corrections were also taken into account.

\section*{Acknowledgements}
We would like to warmly thank Fernando Alday, Alessandro Georgoudis, Tobias Hansen, Andrea Manenti and Alexander Söderberg for collaboration on some of the topics reviewed in this paper and for several discussions. 
We are indebted to Prof. Norma Sanchez for inviting us to contribute with this review article to the Open Access Special Issue ``Women Physicists in Astrophysics, Cosmology and Particle Physics", published in [Universe] (ISSN 2218-1997) and to M. Grana, Y.Lozano, S. Penati and M.Taylor for involving us in this special issue. This work is supported by Knut and Alice Wallenberg Foundation under grant KAW 2016.0129 and by VR grant 2018-04438.

\newpage

\bibliographystyle{unsrt}

\end{document}